\documentclass[fleqn,usenatbib]{mnras}

\usepackage{newtxtext,newtxmath}

\usepackage[T1]{fontenc}

\DeclareRobustCommand{\VAN}[3]{#2}
\let\VANthebibliography\thebibliography
\def\thebibliography{\DeclareRobustCommand{\VAN}[3]{##3}\VANthebibliography}

\usepackage{graphicx}	
\usepackage{amsmath}	
\usepackage{physics}
\usepackage[ruled,vlined]{algorithm2e}
\usepackage{tikz}  
\usepackage{tikz-3dplot}
\usepackage{subcaption}

\title[The MHD-PIC Module in \texttt{Athena++}]{The Magnetohydrodynamic-Particle-In-Cell Module in \texttt{Athena++}: Implementation and Code Tests}

\author[X. Sun \& X.-N. Bai]{
Xiaochen Sun,$^{1}$\thanks{E-mail: sxc18@mails.tsinghua.edu.cn}
Xue-Ning Bai$^{1,2}$\thanks{E-mail: xbai@tsinghua.edu.cn}
\\
$^{1}$Institute for Advanced Study, Tsinghua University, Beijing 100084, China\\
$^{2}$Department of Astronomy, Tsinghua University, Beijing 100084, China
}

\date{Accepted XXX. Received YYY; in original form ZZZ}

\pubyear{2022}

\begin{document}
\label{firstpage}
\pagerange{\pageref{firstpage}--\pageref{lastpage}}
\maketitle

\begin{abstract}
We present a new magnetohydrodynamic-particle-in-cell (MHD-PIC) code integrated into the \texttt{Athena++} framework.
It treats energetic particles as in conventional PIC codes while the rest of thermal plasmas are treated as background fluid described by MHD, thus primarily targeting at multi-scale astrophysical problems involving the kinetic physics of the cosmic-rays (CRs).
The code is optimized toward efficient vectorization in interpolation and particle deposits, with excellent parallel scaling.
The code is also compatible with static/adaptive mesh refinement, with dynamic load balancing to further enhance multi-scale simulations. In addition, we have implemented a compressing/expanding box framework which allows adiabatic driving of CR pressure anisotropy, as well as the $\delta f$ method that can dramatically reduce Poisson noise in problems where distribution function $f$ is only expected to slightly deviate from the background. 
The code performance is demonstrated over a series of benchmark test problems including particle acceleration in non-relativistic parallel shocks. In particular, we reproduce the linear growth of the CR gyro-resonant (streaming and pressure anisotropy) instabilities, under both the periodic and expanding/compressing box setting. We anticipate the code to open up the avenue for a wide range of astrophysical and plasma physics applications.
\end{abstract}



\section{Introduction}
Cosmic rays (CRs) are relativistic charged particles in space, mainly protons, whose energies spans from sub-GeV to $\gtrsim$PeV. It is widely considered that CRs originate from thermal plasma particles and and get accelerated to very high energies through processes such as collisionless shocks \citep[e.g.,][]{1977ICRC....2..273A, 1977DoSSR.234.1306K,1978ApJ...221L..29B, 1978MNRAS.182..443B} and/or magnetic reconnection \citep[e.g.,][]{1998PhPl....5.1599K,2000mare.book.....P, 2022NatRP...4..263J}. Capturing these processes from first principles requires kinetic simulations \citep[e.g.,][]{2011ApJ...726...75S, 2013ApJ...773..158G, 2014ApJ...783...91C, 2014PhRvL.113o5005G, 2014ApJ...783L..21S, 2015ApJ...798L..28C, 2015ApJ...806..167G}.
However, despite of different acceleration physics, there is substantial scale separation between the physical scale that governs particle injection (usually plasma skin depth), the gyration scale that governs wave-particle interaction, and the scale of the global system. 

Another key topic in CR astrophysics is the CR transport and feedback \citep[e.g.,][]{2001RvMP...73.1031F,2017PhPl...24e5402Z, 2017ARA&A..55...59N}. Cosmic rays travel along the interstellar magnetic field after escaping from acceleration sites. Meanwhile, the CRs are scattered by magnetic fluctuations and exchange energy (momentum) with ambient gas. This CR feedback substantially impacts galaxy evolution \citep[e.g.,][]{2015ARA&A..53..199G, 2020MNRAS.496.4221J, 2022MNRAS.tmp.1768H, 2022MNRAS.514..657K}. One major process underlying CR transport and feedback is the CR gyro-resonant instability \citep{1969ApJ...156..445K, 1974ARA&A..12...71W, 1975MNRAS.173..255S, 1975MNRAS.173..245S, 1975MNRAS.172..557S, 2006MNRAS.373.1195L, 2011ApJ...731...35Y, 2020ApJ...890...67Z}. The instability is triggered in the presence of CR streaming or CR pressure anisotropy, and amplifies background magneto-hydrodynamic (MHD) waves via gyro-resonance. The outcome of this instability largely determines the wave amplitudes, modulating the local CR transport coefficients and the efficiency of CR feedback. 
Kinetic simulations \citep{2019ApJ...882....3H, 2019ApJ...876...60B, 2019ICRC...36..279H, 2021ApJ...920..141B, 2022ApJ...928..112B} are essential to capture the underlying micro-physics. Similarly, a huge scale separation makes the simulations highly nontrivial, since the CR gyro-resonant instability operates at CRs' Larmor radii scale (sub-AU for $\sim$GeV particles under typical interstellar medium), which is orders of magnitudes larger than the plasma skin depth, but also orders of magnitudes smaller than the CR mean free paths.

The extreme scale separation challenges the plasma numerical method when conducting simulations to study the kinetic effects of the CRs. The Particle-In-Cell (PIC) method \citep[see][and references therein]{birdsall2018plasma}, the most standard plasma simulation method, must resolve the kinetic physics of both electrons and ions (positrons) \citep[e.g.,][]{1992ApJ...390..454H, buneman1993computer, 2005AIPC..801..345S, 2008PhPl...15e5703B}.  It thus has to overcome tremendous computational difficulties when needing to simultaneously resolve the electron skin depth while to accommodate the CRs. The hybrid-PIC method treats electrons as a (typically) massless fluid and only kinetically iterates ions \citep[e.g.,][]{2002hmst.book.....L, 2003LNP...615..136W, 2007CoPhC.176..419G, 2014JCoPh.259..154K, 2021ApJ...922L..35B, 2022PhPl...29e3902J}. It sacrifices the electron physics to save computing time, but its resolution is still limited by the ion skin depth.
The MHD-PIC method, on the other hand, treats thermal particles as a fluid, described by MHD, while it only treats CRs kinetically as particles following the standard PIC approach. This method emerged early on \citet{1986JCoPh..66..469Z, 2000MNRAS.314...65L}, and was formally introduced relatively recently \citep{2015ApJ...809...55B}. Over the past few years, this method has been implemented in several MHD codes, including \texttt{Athena} \citep{2015ApJ...809...55B}, MPI-AMRVAC \citep{2018MNRAS.473.3394V}, PLUTO \citep{2018ApJ...859...13M}, as well as the particle-based code GIZMO \citep{2022MNRAS.513..282J}. It greatly alleviates the issue of scale separation in CR simulations, and has been gaining popularity in studying CR acceleration in non-relativistic shocks \citep[e.g.,][]{2015ApJ...809...55B, 2018MNRAS.473.3394V} and the CR gyro-resonant instabilities \citep{2018MNRAS.476.2779L, 2019ApJ...876...60B, 2021ApJ...914....3P, 2021ApJ...920..141B, 2022ApJ...928..112B}, and potentially magnetic reconnection \citep{2019PhPl...26a2901D}.

In this paper, we describe the development of the MHD-PIC method on top of the grid-based MHD code \texttt{Athena++} \citep{2020ApJS..249....4S}. \texttt{Athena++} is a rewrite from its predecessor \texttt{Athena} \citep{2008ApJS..178..137S}, which is a higher-order Godunov MHD code written in \texttt{C} and was among the most widely used MHD codes in astrophysical fluid dynamics.
\texttt{Athena++}, written in \texttt{C++}, has a modular design that is much more extensible and versatile, and has already incorporated a wide variety of physics.
It also supports the static/adaptive mesh refinement (SMR/AMR), curvilinear coordinate systems and the flexible grid spacing. 
The single process performance has been substantially improved through efficient vectorization taking the advantage of modern computer architectures. Code execution features dynamical scheduling, which overlaps processes' communication with computation, and achieves an excellent scaling on large distributed memory parallel systems. This is further combined with the load balancing to distribute workload as evenly as possible.
Taking the advantage of the \texttt{Athena++} framework, we design our MHD-PIC module aiming to retains all these merits. In particular, 
our MHD-PIC module is implemented as an independent part in \texttt{Athena++}, compatible with mesh refinement and dynamical load balancing.
The combination of adaptive mesh refinement (AMR) and dynamical load balancing benefits many potential applications such as CR acceleration in shocks \citep[e.g.,][]{2018MNRAS.473.3394V} and/or magnetic reconnection. Meanwhile, we optimize the MHD-PIC module by redesigning the data layouts to improve the vectorization efficiency during interpolation and deposits.

We further implement additional technical methods in our MHD-PIC code, mostly for studying the gyro-resonant instability. The $\delta f$ method \citep[e.g.][]{1993PhFlB...5...77P, 1994PhPl....1..863H, 1995JCoPh.119..283D, 2014JCoPh.259..154K, 2019ApJ...876...60B} only evolves the CRs deviating from an equilibrium distribution and can suppress the statistic noise associated with the representation of the CR ensemble by a finite number of discrete Lagrangian particles. Such method plays a key role in the gyro-resonant instability simulations where a small deviation of the CR distribution from isotropy effectively amplifies/damps the MHD waves. In addition, one flavor of the gyro-resonant instability, the CR pressure anisotropy instability \citep[CRPAI, ][]{2006MNRAS.373.1195L, 2011ApJ...731...35Y, 2020ApJ...890...67Z}, can be triggered by background
expansion/compression in astronomical environments. 
To mimic the reality, we formulate and implement the ``expanding box'' framework \citep{1993PhRvL..70.2190G, 2015ApJ...800...88S, 2021ApJ...922L..35B} for our MHD-PIC code which can continuously drive CR pressure anisotropy to enable detailed studies of the CRPAI.

We examine our implementation with several benchmark test problems. Standard tests include the CR gyration test and the Bell instability \citep{2004MNRAS.353..550B}.
We next set up additional test simulations that are more closely connected to real astrophysical applications.
We consider CR acceleration in parallel shocks,
where we enable AMR capability near shock fronts and examine the load balancing performance.
Enabling the $\delta f$ method, we further measure the linear growth rates of the CR gyro-resonant instabilities and compare the results with analytical dispersion relations. We provide additional tests for the ``expanding box" framework, and present preliminary studies of the CRPAI under this framework.

\section{MHD-PIC equations and numerical integration methods}
\label{sec::mhd_pic}

The MHD-PIC equations contain two parts, the equation of motion for the CRs and the MHD equations for the background gas. The CRs are represented by individual super particles, a collection of CR particles sharing similar phase space positions, and integrated according to the Lorentz force. The background gas is described by ideal MHD equations, with the electric field largely determined by the frozen-in condition. In this section, we briefly review the MHD-PIC equations and the numerical algorithm for integrating them \citep[see details in][]{2015ApJ...809...55B, 2018MNRAS.473.3394V, 2018ApJ...859...13M}.

\subsection{CRs' equation of motion}
\label{sec::pic_integrate}

For one CR (super-)particle $k$, let us denote its phase-space information by its position $\boldsymbol{x}_k$ and mass-normalized momentum $\left(\boldsymbol{p} / m\right)_k$. They evolve according to
\begin{align}
	\frac{\dd \boldsymbol{x}_k}{\dd t} &= \boldsymbol{v}_k, \notag \\ 
	\frac{\dd \left(\boldsymbol{p} / m\right)_k}{\dd t} &= \left(\frac{q}{mc}\right)_\alpha \left(c \boldsymbol{E} + \boldsymbol{v}_k \times \boldsymbol{B} \right), \label{equ::pic_dynamic}
\end{align}
where $\left(q /\left(mc\right)\right)_\alpha$ represents the particle charge-to-mass ratio, and the subscript $\alpha$ denotes the particle species. It is related to the CR cyclotron frequency by $\Omega_{\alpha} = \left(q /\left(mc\right)\right)_\alpha |\boldsymbol{B}|$. Note that the speed of light $c$ can always be absorbed to the (non-relativistic) MHD equations and through the $\left(q /\left(mc\right)\right)$ factor in the CR equation of motion. The Lorentz force depends on the electric and magnetic field where particle $k$ is located, $\boldsymbol{E}$ and $\boldsymbol{B}$, which will be discussed later in this section.

It is convenient to define an artificial speed of light $\mathbb{C}$, so that
\begin{align}
	\gamma_k &= \sqrt{1 + \left(\boldsymbol{p} / m\right)_k^2 / \mathbb{C}^2}, \quad
	\boldsymbol{v}_k = \frac{\left(\boldsymbol{p} / m\right)_k}{\gamma_k}, \notag \\ 
	\varepsilon_k &= \frac{\left(\boldsymbol{p} / m\right)_k^2}{1 + \gamma_k} = \left(\gamma_k - 1\right) \mathbb{C}^2,
\end{align}
where $\gamma_k$, $\boldsymbol{v}_k$ and $\varepsilon_k$ are the Lorentz factor, particle velocity, and the specific particle kinetic energy $\varepsilon_k$. 
There is some freedom in choosing $\mathbb{C}$ as long as sufficient scale separation is respected (which usually favors large $\mathbb{C}$) while optimizing the computational cost ($\mathbb{C}$ is not too large). This $\mathbb{C}$ is then considered to be the real speed of light for the CRs in the simulations, and when applying to physical systems, the results can be rescaled accordingly.\footnote{Alternatively, given the physical system, one may adopt the reduced speed of light formulation by \citep{2022MNRAS.516.5143J} to save computational cost, which requires applying a number of scaling factors to the formulation and it ensures the correctness of the solutions in steady state.} This is also similar to the routine practice in standard PIC simulations.

\subsection{Equations of gas dynamics}
\label{sec::mhd_integrate}
When including CR backreaction, the equations of gas dynamics read \citep{2015ApJ...809...55B}
\begin{align}
	\partial_t \rho + \nabla \cdot \left(\rho \boldsymbol{u}\right) = 0, \label{equ::mhd_mass} \\ 
	\partial_t \left(\rho \boldsymbol{u}\right) + \nabla \cdot \left(\rho \boldsymbol{u^T} \boldsymbol{u} - \boldsymbol{B^T} \boldsymbol{B}\right) + \nabla \left(P + B^2 / 2\right) \notag \\ = -\left(\frac{\rho_{CR}}{c} c\boldsymbol{E} + \frac{\boldsymbol{j}_{CR}}{c} \times \boldsymbol{B} \right), \label{equ::mhd_momentum} \\ 
	\partial_t \mathcal{E} + \nabla \cdot \left[\left(\mathcal{E} + P + B^2 / 2\right) \boldsymbol{u} - \left(\boldsymbol{u} \cdot \boldsymbol{B}\right) \boldsymbol{B} \right] = - \frac{\boldsymbol{j}_{CR}}{c} \cdot \left(c\boldsymbol{E}\right). \label{equ::mhd_energy}
\end{align}
The gas energy density $\mathcal{E}$ is given by
\begin{equation*}
	\mathcal{E} = \frac{P}{\Gamma - 1} + \frac{1}{2} \rho u^2 + \frac{1}{2} B^2,
\end{equation*}
where $\Gamma$ is the gas adiabatic index. All other variables above have their usual meanings,
with $\rho$ for gas density, $P$ for gas pressure, and ${\boldsymbol u}$ for gas velocity.
The units for magnetic field has magnetic permeability being one, so that factors of $1/\sqrt{4\pi}$ are readily absorbed.
The CRs additionally exert the Lorentz force with the corresponding energy deposition, associated with CR charge density and current density given by
\begin{align*}
    \frac{\rho_{CR}}{c} \equiv \sum_{\alpha} \left(\frac{q}{c}\right)_\alpha \int f_\alpha \dd^3 \boldsymbol{p} =& \sum_{\alpha} \left(\frac{q}{mc}\right)_\alpha m_\alpha \int f_\alpha \dd^3 \boldsymbol{p},  \\
	\frac{\boldsymbol{j}_{CR}}{c} \equiv \sum_{\alpha} \left(\frac{q}{c}\right)_\alpha \int \boldsymbol{v} f_\alpha \dd^3 \boldsymbol{p} =& \sum_{\alpha} \left(\frac{q}{mc}\right)_\alpha m_\alpha \int \boldsymbol{v} f_\alpha \dd^3 \boldsymbol{p} \notag .
\end{align*}
Here $f_\alpha\left(t, \boldsymbol{x}, \boldsymbol{p} \right)$ is the CRs' phase distribution function for species $\alpha$ and $m_\alpha$ is the real CR particle rest mass. As a super-particle, one simulation particle represents a swarm of real CR particles.
The mass of a simulation particle $m_\alpha^{\text{sim}}$ is chosen so that its product with particle number density $n_\alpha^{\text{sim}}$, or the CR mass density $m_\alpha^{\text{sim}} n_\alpha^{\text{sim}}=m_\alpha n_\alpha \equiv  m_\alpha \int f_\alpha \dd^3 \boldsymbol{p}$ conform to the desired physical values.

In this work, we adopt the ideal MHD induction equation with the associated electric field
\begin{equation}
\partial_t \boldsymbol{B} =-\nabla \times (c\boldsymbol{E}),\qquad
	c\boldsymbol{E} = -\boldsymbol{u} \times \boldsymbol{B}. \label{equ::frozen_in}
\end{equation}
In doing so, we neglect the CR-induced Hall term \citep{2015ApJ...809...55B}, which is important when the product $j_{\rm CR}/c\gtrsim (q/mc)\rho U_A$. This condition is likely satisfied only in the vicinity of strong shocks, yet this term shows very limited impact.

\subsection{Implementation of the basic version in \texttt{Athena++}}
\label{sec::num_mhd-pic}
Here, we briefly outline the basic numerical integration methods for uniform grid. Additional extensions will be further addressed in the next two sections.

We follow the basic design in \texttt{Athena++}, which divides the computational domain into sub-volumes named ``MeshBlocks'' as independent computational unit that can be distributed to individual processors for parallel computing. 
Information of particles that belong to each MeshBlock is stored within the MeshBlock, so as to share the information with the MHD grid data. Within individual MeshBlocks, the particle data are stored as structures of arrays, including arrays of their physical location $\boldsymbol{x}$ and the mass-normalized momentum $\boldsymbol{p} / m$. In the case of $\delta f$ method (see Section \ref{ssec:deltaf}), the value of the initial distribution function $f\left(t=0, \boldsymbol{x}\left(t=0\right), \boldsymbol{p}/m\left(t=0\right)\right)$ is also stored. Particles that cross MeshBlock boundaries are removed, while particles arriving from neighboring MeshBlocks are attached to the end of the particle arrays.

We apply the van Leer time integrator \citep{2009NewA...14..139S} to solve the MHD equations \footnote{It is possible to use other second-order integrators \citep{2018ApJ...859...13M}. Here we consider the van Leer integrator for its better stability and that it is more straightforward to couple with the Boris pusher. The MHD-PIC module implemented in \citet{2015ApJ...809...55B} was on top of the CTU integrator, where the overall numerical scheme is similar to that of the van Leer integrator.} and the Boris pusher \citep{boris1972proceedings} to integrate CR particles. The van Leer time integrator and the Boris pusher employ two stages to integrate for a full time step $\Delta t$, beginning from time $t^{ini}$ and ending at time $t^{fin}$, which achieves second-order temporal accuracy. 

We here list the numerical procedures associated with the MHD-PIC part, while the details about the MHD integrator have been introduced in \citet{2020ApJS..249....4S}. The first stage is to predict the midpoint states (e.g. gas velocity, magnetic field, CRs' positions at $t^{mid}= t^{ini}+\Delta t/2$) based on the results from the previous step:
\begin{itemize}
    \item Stage 1.1: Compute CR particles' weights over neighboring cells according to the initial position $\boldsymbol{x}^{ini}$.
    \item Stage 1.2: Deposit $\rho_\text{CR} / c$ \& $\boldsymbol{j}_\text{CR}/ c$ from the CR particles.
    \item Stage 1.3: Move all CR particles forward by $\Delta t/2$ based on their initial velocities, from $\boldsymbol{x}^{ini}$ to $\boldsymbol{x}^{mid}$.
    \item Stage 1.4: Integrate MHD gas from $t^{ini}$ to $t^{mid}$ by means of the Riemann solver and constrained transport \citep{2005JCoPh.205..509G, 2008JCoPh.227.4123G, 2020ApJS..249....4S}. The momentum and energy source terms from the CRs are added to the gas as source terms.
    \item Stage 1.5: Apply boundary conditions to both MHD gas and CR particles.
\end{itemize}
The second stage updates for a full time step, based on the the midpoint states from the first stage:
\begin{itemize}
    \item Stage 2.1: Compute CR particles' weights over neighboring cells according to the midpoint position $\boldsymbol{x}^{mid}$.
    \item Stage 2.2: For each CR particle $k$, interpolate $\boldsymbol{u}_k$ \& $\boldsymbol{B}_k$ from the gas to the particle location $\boldsymbol{x}_k^{mid}$, and compute $\boldsymbol{E}_k$.
    \item Stage 2.3: Integrate the CR momentum with the Boris pusher from $t^{ini}$ to $t^{fin}$ based on the interpolated electro-magnetic field, and record the momentum change $\Delta \left(\frac{\boldsymbol{p}}{m}\right)_k$ \& energy change $\Delta \varepsilon_k$ for each CR particle $k$. 
    \item Stage 2.4: Deposit the CR momentum gain $\int \Delta \frac{\boldsymbol{p}}{m} f \dd^3 \boldsymbol{p}$ \& the CR energy gain $\int \Delta \varepsilon f \dd^3 \boldsymbol{p}$ to the gas.
    \item Stage 2.5: Move all CR particles forward by $\Delta t/2$ based on their final velocities, from $\boldsymbol{x}_k^{mid}$ to $\boldsymbol{x}_k^{fin}$.
    \item Stage 2.6: Integrate the MHD gas from $t^{ini}$ to $t^{fin}$, and then subtract the CR momentum and energy gain from the gas.
    \item Stage 2.7: Apply boundary conditions to both the MHD gas and CR particles.
\end{itemize}
The interpolation and deposit in the above procedures follows the the standard triangular-shaped cloud (TSC) scheme \citep{birdsall2004plasma}. To ensure $\boldsymbol{B} \cdot \boldsymbol{E}=0$, we choose the gas velocity $\boldsymbol{u}_k$ and the magnetic field $\boldsymbol{B}_k$ as the interpolated variables to position $\boldsymbol{x}_k^{mid}$, so that the electric field $\boldsymbol{E}_k$ computed via the frozen-in condition (Equation~\ref{equ::frozen_in}) is always perpendicular to $\boldsymbol{B}_k$. Meanwhile, note that we deposit CRs' charge and current density for the CR feedback without particle integration in the first stage. Conservation of total momentum and energy is ensured in the second stage where CRs' momentum and energy difference are deposited.
Also note that for certain situations, when CR backreaction is negligible, it is natural to neglect CR feedback in Stage~1.1, 1.2~\&~2.4 and treat the CRs as test particles.

Similar to \citet{2015ApJ...809...55B}, the integration timestep $\Delta t$ is constrained from the following three conditions,
\begin{itemize}
    \item The standard Courant-Friedrichs-Lewy stability condition of the van Leer integrator \citep{2009NewA...14..139S, 2020ApJS..249....4S}.
    \item The CR particles cannot cross more than $N_\text{max}=2$ cells in $\Delta t$.
    \item The CR particles should not rotate by more than $\theta_\text{max}$ radian in $\Delta t$.
\end{itemize}
In general, we recommend choosing $\theta_\text{max}\lesssim0.3$ to ensure sufficient accuracy in particle orbital integration.

The basic integration applies to Cartesian and uniform grid. Extension to include mesh refinement will be discussed in Section~\ref{sec::refinement}. Generalization of the Boris pusher to curvi-linear coordinates can be achieved by applying transformation to Cartesian coordinates. While the TSC scheme for interpolation/deposits can be applied to curvi-linear coordinates and non-uniform grids which is supported in the code, we caution that it would fail to ensure that a spatially uniform distribution of particle deposit uniformly to the grid.
On the other hand, we anticipate that mesh refinement should serve the need of most applications that requires adaptive resolution, we refrain from further discussions on non-uniform grids or curvi-linear coordinates.

\section{The Expanding Box Extension}
One of the scientific applications for our code is to study the CRPAI (see Section~\ref{sec::test_anisotropy}), where the CR anisotropy is naturally driven by the expansion or compression of the background flow. For this purpose, we further develop an expanding-box module into the standard MHD-PIC framework.
The expanding box approach involves formulating the equations in co-moving coordinates on top of an expanding or contracting background, which has been widely applied in the simulations of the solar wind \citep[e.g.,][]{1993PhRvL..70.2190G,2020ApJ...888...68S,2020ApJ...891L...2S}. Recently, this approach has also been employed to study anisotropy-driven plasma kinetic instabilities \citep{2015ApJ...800...88S,2015ApJ...800...89S,2021ApJ...922L..35B}, where background expansion or contraction keeps driving pressure anisotropy in collisionless particles associated with the conservation of magnetic moments. Here we apply this formulation to the MHD-PIC framework.
Our formulation is the most general in the sense that we allow the expansion/contraction in the three directions be independent of each other. We first derive the expanding box formulation below, followed by a description of its implementation.

Analogous to the case of the expanding universe, we derive the expanding box formulas in comoving coordinates, using an anisotropic Friedmann-Robertson-Walker (FRW) metric in the general relativistic framework. In general, the box expansion rate $a_i(t)$ can differ in each direction, $a_i(t) \dd x'^i =\dd x^i$, where $\boldsymbol{x}$ and $\boldsymbol{x}'$ denote the position in the lab frame and the comoving frame, respectively. The geodesic distance is $c^2 \dd^2 \tau =\eta_{\mu \nu} \dd x^\mu \dd x^\nu = g_{\mu \nu} \dd x'^\mu \dd x'^\nu$, with the expanding box metric $\mathbf{g} = diag\left(-1, a_1^2(t), a_2^2(t), a_3^2(t)\right)$, where $\tau$ is the proper time. In this subsection, Greek letters are time-space indices which go from 0 to 3, while $i, j, k$ are space indices which go from 1 to 3. Einstein summation is solely applied to Greek letters. We generalize the standard MHD-PIC equations to covariant form of the comoving frame quantities in this FRW-like geometry. The general-relativistic MHD equations with CR backreaction consist of the gas rest mass conservation, the energy-momentum equation, the Faraday–Gauss equation and the frozen-in theorem:
\begin{align}
    \left(\rho U'^\mu \right)_{;\mu} = 0, \label{equ::mass_mhd_gr} \\
	T'^{\mu \nu}_{;\nu} = - g^{\mu \nu} F'_{\nu \kappa}  J'^{\kappa}, \label{equ::energy_momentum_mhd_gr} \\
    F'_{\nu \kappa ; \mu} + F'_{\kappa \mu ;\nu} + F'_{\mu \nu; \kappa}=0, \\
    U'^\mu F'_{\mu \nu} = 0, \label{equ::frozen_in_gr} 
\end{align}
where
\begin{align*}
    F'_{ij} \equiv [ijk] l  B_k / a_k, \\
    T'^{\mu \nu} \equiv \frac{P }{\Gamma - 1} \frac{ U'^\mu U'^\nu}{c^2} + T'^{\mu \nu}_\text{HD} + T'^{\mu \nu}_\text{EM}, \\
    T'^{\mu \nu}_\text{HD} \equiv \left(\rho + P/c^2\right)U'^\mu U'^\nu + P g^{\mu \nu},\\
    T'^{\mu \nu}_\text{EM} \equiv F'^{\mu \alpha} F'^\nu_{\alpha} - \frac{1}{4} g^{\mu \nu} F'^{\alpha \beta} F'_{\alpha \beta}.
\end{align*}
The equation of motion for a charged particle in a curved space-time is:
\begin{equation}
	\frac{\dd^2 x'^\mu}{\dd \tau^2} = \frac{q}{mc} g^{\mu \nu} F'_{\nu \kappa} \frac{\dd x'^\kappa}{\dd \tau}, \label{equ::pic_dynamic_gr}
\end{equation}
Here $U'^\mu$, $F'_{\mu \nu}$, $T'^{\mu \nu}$ and $J'^\mu$ are the gas four-velocity, the electromagnetic tensor, the MHD energy-momentum tensor, and CRs' charge-current density in the co-moving frame, respectively. The determinant of the metric $l(t)$ reflects the volume expansion rate,
\begin{equation}
    l(t) \equiv \prod_i a_i = \sqrt{- \det (\mathbf{g})}.
\end{equation}
The subscript $_{; \mu}$ represents the covariant derivative with respect to $\mu$ and $[ijk]$ is the anti-symmetric tensor. The electric field, the time-like component in $F'_{\mu \nu}$, is determined from the frozen-in condition (Equation~\ref{equ::frozen_in_gr}). The energy-momentum tensor comprises of contributions from the electromagnetic field $T'^{\mu \nu}_\text{EM}$, hydrodynamic energy-momentum tensor $T'^{\mu \nu}_\text{HD}$ and the thermal energy density $P / \left(\Gamma - 1\right)$ (Equation~\ref{equ::energy_momentum_mhd_gr}).

The MHD-PIC equations in expanding box can be substantially simplified in the non-relativistic limit that usually applies with $U'^i \approx u_i / a_i \ll c$, by substituting the comoving frame quantities (primed) back to the lab frame quantities (unprimed) and expressing the spatial derivatives in the comoving frame. Equation~\ref{equ::mass_mhd_gr}~to~\ref{equ::frozen_in_gr} are reduced to:
\begin{align}
    \partial_t \rho + \nabla' \cdot \left(\rho \boldsymbol{u}\right) = - \rho \frac{\partial_t l}{l}  , \label{equ::expand_mass_mhd} 
\end{align}
\begin{align}
    \partial_t \left(\rho \boldsymbol{u}\right) + &\nabla' \cdot \left(\rho \boldsymbol{u^T} \boldsymbol{u} - \boldsymbol{B^T} \boldsymbol{B}\right) + \nabla' \left(P + B^2 / 2\right) \notag \\ =& - \rho \boldsymbol{u} \frac{\partial_t l}{l} - \rho \boldsymbol{u} \cdot \mathbb{D}  -\left(\frac{\rho_{CR}}{c} c\boldsymbol{E} + \frac{\boldsymbol{j}_{CR}}{c} \times \boldsymbol{B} \right), \label{equ::expand_momentum_mhd}
\end{align}
\begin{align}
    &\partial_t \mathcal{E} + \nabla' \cdot \left[\left(\mathcal{E} + P + B^2 / 2\right) \boldsymbol{u} - \left(\boldsymbol{u} \cdot \boldsymbol{B}\right) \boldsymbol{B} \right] \notag   \\=& -\frac{\partial_t l}{l} \left(\mathcal{E} + P\right)  - \rho \boldsymbol{u^T} \cdot \mathbb{D} \cdot \boldsymbol{u} + \boldsymbol{B^T} \cdot \mathbb{D} \cdot \boldsymbol{B} = - \frac{\boldsymbol{j}_{CR}}{c} \cdot \left(c\boldsymbol{E}\right), \label{equ::expand_energy_mhd}
\end{align}
\begin{align}
    \nabla' \cdot \boldsymbol{B} = 0, \label{equ::expand_gauss}
\end{align}
\begin{align}
    \partial_{t} \boldsymbol{B}  - \nabla' \cross \left(\boldsymbol{u}\times\boldsymbol{B}\right) = - \frac{\partial_t l}{l}\boldsymbol{B} + \mathbb{D} \cdot \boldsymbol{B}, \label{equ::expand_field} 
\end{align}
where
\begin{align}
    c\boldsymbol{E} = - \boldsymbol{u} \times \boldsymbol{B},
\end{align}
\begin{align}
    \mathbb{D} = diag\left(\frac{\partial_t a_1}{a_1}, \frac{\partial_t a_2}{a_2},\frac{\partial_t a_3}{a_3}\right),\notag 
\end{align}
and the gradient operator $\nabla'$ is equivalent to the lab frame one, but expressed in the co-moving coordinates,
\begin{align}
    \nabla' = \left(\frac{\partial}{a_1 \partial x'_1} ,\frac{\partial}{a_2 \partial x'_2}, \frac{\partial}{a_3\partial x'_3} \right). \notag
\end{align}
The CR source terms in Equation~\ref{equ::expand_momentum_mhd}~\&~\ref{equ::expand_energy_mhd} are also expressed with the lab frame current density, which stays in the original form. It is straightforward to explicitly evolve the CR momentum in the lab frame, $\left(p_i / m\right)= a_i \dd x'^i / \dd \tau$, and the CR equation of motion (Equation~\ref{equ::pic_dynamic_gr}) becomes:
\begin{align}
    \mathbb{A} \cdot \frac{\dd \boldsymbol{x}'}{\dd t} = \boldsymbol{v}, 
\end{align}
\begin{align}
    \frac{\dd \left(\boldsymbol{p} / m\right)}{\dd t} + \mathbb{D} \cdot \frac{\boldsymbol{p}}{m} = \frac{q}{mc} \left(c \boldsymbol{E} + \boldsymbol{v} \times \boldsymbol{B} \right), 
\end{align}
\begin{align}
    \mathbb{A} = diag \left(a_1, a_2, a_3\right).\notag
\end{align}

Specifically, when only $x_2-$ and $x_3-$ directions expand at the same rate, $a_1 = 1, a_2=a_3=a(t)$, above equations reduce to the cases in \citet{1993PhRvL..70.2190G} and \citet{2015ApJ...800...88S}. Compared to the formulas in Section~\ref{sec::pic_integrate}~\&~\ref{sec::mhd_integrate}, due to the lab frame quantities we have chosen, the dynamic equations are mostly similar the conventional ones in the flat spacetime. The additional terms in above equations clearly reflect the expansion effect. For a uniform expanding box without source terms (CR backreaction and the Lorentz force acting on the CRs), the contribution from these expansion-driven terms can be recast into,
\begin{align}
    &\partial_t \left(l \rho \right)= 0, \quad
    &\partial_t \left(\boldsymbol{u} \cdot \mathbb{A}\right) = 0,\quad
    &\partial_t \left(\left(\boldsymbol{p} / m\right) \cdot \mathbb{A}\right) = 0, \notag \\
    &\partial_t \left( l^\Gamma P\right) = 0,\quad
	&\partial_t \left( l B_i / a_i \right) = 0, \label{equ::expansion_driven}
\end{align}
which express the mass ($\rho \dd^3 \boldsymbol{x}$) conservation law in the lab frame, the anisotropic stretching for MHD gas and CRs \citep{1993PhRvL..70.2190G}, the adiabatic polytropic process ($P \left(\dd^3 \boldsymbol{x}\right)^\Gamma$) and the magnetic flux ($\left[ijk\right]B_i \dd x^j \dd x^k$) conservation in the lab frame.

We implement the expansion-driven terms at the end of each stage by following the relation given by Equation~\ref{equ::expansion_driven}. For instance, consider the update of the $i$th component of magnetic field $B_i$ at the stage $n$, from time $t^1$ to $t^2$. Following the standard constrained transport, $B_i$ is first updated from $B_i^{1}$ to $B_{i, flat}^{2}$, where ``flat" indicates flat spacetime (no expansion). The expanding box source term can be implemented via a rescaling,
\begin{equation*}
	B_i^{2} = \frac{l\left(t^{1} \right) / a_i \left(t^{1}\right)}{l\left(t^{2} \right) / a_i \left(t^{2}  \right)} \times B_{i, flat}^{2}.
\end{equation*}
The term $l/a_i$ represents the expansion rate for a face area normal to direction $i$ and the multiplied factor is the area change ratio over the integration timestep. The gas density and gas momentum are similarly rescaled with the ratio of $l$ and $l\mathbb{A}$, respectively, at the end of each stage. The treatment of the gas energy density is more complex but in the same way, where the gas thermal energy density, the gas kinetic energy density and magnetic field energy density are multiplied by different factors. This method ensures the stability and accuracy in high-order time integrators and conserve the magnetic field divergence. For the CR expansion-driven term, similar to the method in \cite{2015ApJ...800...88S}, we let the expansion-driven term $\mathbb{D} \cdot \boldsymbol{p} / m$ act on particles for a half timestep before and after the Boris pusher.
We provide benchmark test problems for each of the MHD and CR parts to validate our implementations of the expanding box module in Appendix~\ref{app::test_expanding}.

\section{Code Optimization and Advanced Features}

In this section, on top of the basic MHD-PIC integration method described in Section \ref{sec::num_mhd-pic}, we describe techniques to optimize and enhance code performance especially with respect to interpolation and deposits (Section \ref{sec::vec}), parallelization and load balancing (Section \ref{sec::load}). We further discuss additional extensions accommodate adaptive and static mesh refinement (Section \ref{sec::refinement}), as well as the $\delta f$ method (Section \ref{ssec:deltaf}).

\subsection{Vectorization, intermediate arrays and sorting}
\label{sec::vec}
The grid layout in the MHD part makes the \texttt{Athena++} code very efficient through vectorization.
However, while the particle part of the code is expected take most of the computational time for typical MHD-PIC applications, it is not easily vectorized. A main bottleneck lies in the interpolation and deposit steps, where each particle must access its neighboring multidimensional grid points that are not stored contiguously in memory. Here we discuss our strategies to enhance code performance that enables relatively efficient vectorization in the particle module.

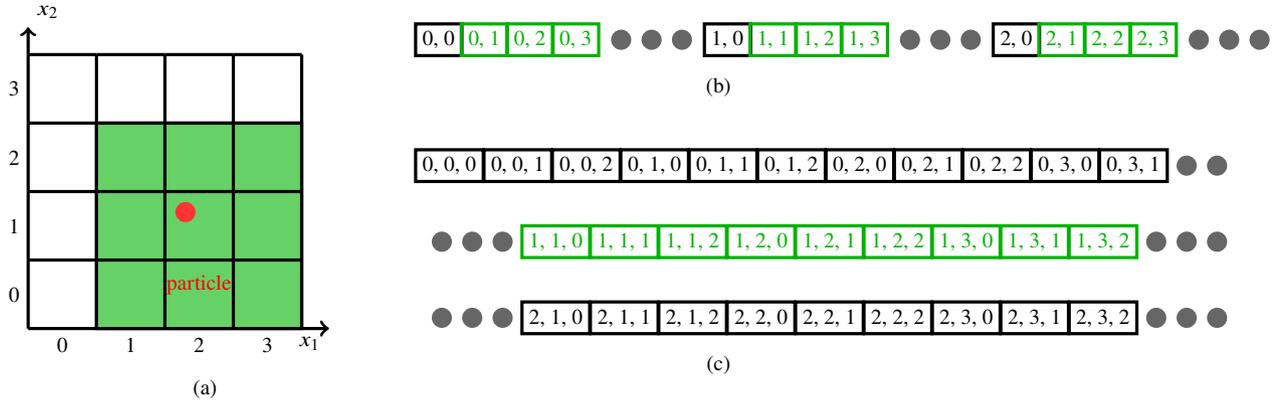
\begin{figure*}
	\centering
	\begin{subfigure}[]{.3\textwidth}
		\begin{tikzpicture}[scale=0.9,very thick]
        \draw[fill=green!70!black, fill opacity=0.6] (1,0) -- (4,0) -- (4,3) -- (1,3) -- cycle;
		\foreach \x in {0,1,2,3,4}
		\foreach \y in {0,1,2,3,4}{
		\ifthenelse{  \lengthtest{\x pt < 4pt}  }{
		\draw (\x,\y) -- (\x+1,\y);
		}{}
		\ifthenelse{  \lengthtest{\y pt < 4pt}  }{
		\draw (\x,\y) -- (\x,\y+1);
		}{}
		}
        \foreach \x in {0,1,2,3}{
        \draw (\x+0.5,0) node[below] {\x};
        }
        \foreach \y in {0,1,2,3}{
        \draw (0,\y+0.5) node[left] {\y};
        }
        
		\draw (2.3,1.7) node[circle,fill,color = red!80] {}; 
		\draw[->](4,0)--(4.4,0) node[below left]{$x_1$};
		\draw[->](0,4)--(0,4.4) node[above right]{$x_2$};

		\draw[color=red] (2.5,0.9) node[below] {particle};
		\end{tikzpicture}
		\caption{}
		\label{fig::par_3d}
	\end{subfigure}
    \begin{minipage}[]{.67\textwidth}
	\begin{subfigure}[]{.67\textwidth}
        \centering
		\begin{tikzpicture}[scale=1.0,very thick]
            \foreach \x in {0,1,2,3}{
                \ifthenelse{  \lengthtest{\x pt > 0pt}  }{
		      \draw (\x * 0.6  ,1 -0)  node[rectangle, draw=green!70!black, text = green!70!black] {0, \x};
		      }{\draw (\x * 0.6 ,1 -0)  node[rectangle, draw] {0, \x};
                }
                \ifthenelse{  \lengthtest{\x pt > 0pt}  }{
		      \draw (\x * 0.6 + 3.8 ,1 -0)  node[rectangle, draw=green!70!black, text = green!70!black] {1, \x};
		      }{\draw (\x * 0.6 + 3.8,1 -0)  node[rectangle, draw] {1, \x};
                }
                \ifthenelse{  \lengthtest{\x pt > 0pt}  }{
		      \draw (\x * 0.6 + 7.6 ,1)  node[rectangle, draw=green!70!black, text = green!70!black] {2, \x};
		      }{\draw (\x * 0.6 + 7.6,1)  node[rectangle, draw] {2, \x};
                }
    		}
            \foreach \x in {0,1,2}{
            \draw (\x * 0.4 + 2.4,1) node[circle,fill,color = black!60] {};
            \draw (\x * 0.4 + 6.2,1) node[circle,fill,color = black!60] {};
            \draw (\x * 0.4 + 10.0,1) node[circle,fill,color = black!60] {};
            }
        \end{tikzpicture}
		\caption{}
		\label{fig::cc_org}
	\end{subfigure}\\ \vspace{2.0em} \\
	\begin{subfigure}[]{.67\textwidth}
		\centering
		\begin{tikzpicture}[scale=1.0,very thick]
            \draw (0.9 * 0, 2)  node[rectangle, draw] {0, 0, 0};
            \draw (0.9 * 1, 2)  node[rectangle, draw] {0, 0, 1};
            \draw (0.9 * 2, 2)  node[rectangle, draw] {0, 0, 2};
            \draw (0.9 * 3, 2)  node[rectangle, draw] {0, 1, 0};
            \draw (0.9 * 4, 2)  node[rectangle, draw] {0, 1, 1};
            \draw (0.9 * 5, 2)  node[rectangle, draw] {0, 1, 2};
            \draw (0.9 * 6, 2)  node[rectangle, draw] {0, 2, 0};
            \draw (0.9 * 7, 2)  node[rectangle, draw] {0, 2, 1};
            \draw (0.9 * 8, 2)  node[rectangle, draw] {0, 2, 2};
            \draw (0.9 * 9, 2)  node[rectangle, draw] {0, 3, 0};
            \draw (0.9 * 10, 2)  node[rectangle, draw] {0, 3, 1};
            \foreach \y in {0,1}
            \foreach \x in {0,1,2}{  
            \draw (\x * 0.4 - 0.1,1-\y) node[circle,fill,color = black!60] {};
            \draw (\x * 0.4 + 9.3,1-\y) node[circle,fill,color = black!60] {};
            }
            \foreach \x in {0,1}{  
            \draw (\x * 0.4 + 9.7,2) node[circle,fill,color = black!60] {};
            }
            \draw (0.9 * 0 + 1.4, 1)  node[rectangle, draw=green!70!black, text = green!70!black] {1, 1, 0};
            \draw (0.9 * 1 + 1.4, 1)  node[rectangle, draw=green!70!black, text = green!70!black] {1, 1, 1};
            \draw (0.9 * 2 + 1.4, 1)  node[rectangle, draw=green!70!black, text = green!70!black] {1, 1, 2};
            \draw (0.9 * 3 + 1.4, 1)  node[rectangle, draw=green!70!black, text = green!70!black] {1, 2, 0};
            \draw (0.9 * 4 + 1.4, 1)  node[rectangle, draw=green!70!black, text = green!70!black] {1, 2, 1};
            \draw (0.9 * 5 + 1.4, 1)  node[rectangle, draw=green!70!black, text = green!70!black] {1, 2, 2};
            \draw (0.9 * 6 + 1.4, 1)  node[rectangle, draw=green!70!black, text = green!70!black] {1, 3, 0};
            \draw (0.9 * 7 + 1.4, 1)  node[rectangle, draw=green!70!black, text = green!70!black] {1, 3, 1};
            \draw (0.9 * 8 + 1.4, 1)  node[rectangle, draw=green!70!black, text = green!70!black] {1, 3, 2};
            \draw (0.9 * 0 + 1.4, 0)  node[rectangle, draw] {2, 1, 0};
            \draw (0.9 * 1 + 1.4, 0)  node[rectangle, draw] {2, 1, 1};
            \draw (0.9 * 2 + 1.4, 0)  node[rectangle, draw] {2, 1, 2};
            \draw (0.9 * 3 + 1.4, 0)  node[rectangle, draw] {2, 2, 0};
            \draw (0.9 * 4 + 1.4, 0)  node[rectangle, draw] {2, 2, 1};
            \draw (0.9 * 5 + 1.4, 0)  node[rectangle, draw] {2, 2, 2};
            \draw (0.9 * 6 + 1.4, 0)  node[rectangle, draw] {2, 3, 0};
            \draw (0.9 * 7 + 1.4, 0)  node[rectangle, draw] {2, 3, 1};
            \draw (0.9 * 8 + 1.4, 0)  node[rectangle, draw] {2, 3, 2};  
        \end{tikzpicture}
		\caption{}
		\label{fig::interm}
	\end{subfigure}
    \end{minipage}
	\caption{Illustration of the intermediate array in 2D. (\ref{fig::par_3d}): One particle (red dot) located at $(x_1=1.8,x_2=1.2)$ in the MHD grid,
        and the green shaded region corresponds to the grids points with which the red particle exchanges information from (interpolation and deposits). (\ref{fig::cc_org}): Layout of the original MHD data in the memory. The numbers correspond to the spatial indices $(j,i)$ for $(x_2,x_1)$ and the green boxes correspond to the 
    cells in the shaded region in Figure~\ref{fig::par_3d}. The grey dots represents data from other cells stored in between. (\ref{fig::interm}): Data layout for the intermediate array. The outer two indices (from the left) are the original grid indices and the innermost dimension corresponds to the relative position to the neighbours in the $y$ direction. Also, the green data are the ones that the red particle in Figure~\ref{fig::par_3d} interpolates from or deposits to.}
\end{figure*}

To illustrate our strategy, we consider particle interpolation/deposit on a 2D grid in Figure~\ref{fig::par_3d}.
A particle needs data from 9 neighboring cells in the TSC scheme. The corresponding MHD data, whose inner index is along the $x_1-$direction, are separated into three discontinuous sections in the computer memory, as shown in Figure~\ref{fig::cc_org}. A naive implementation of interpolation/deposit would likely encounter multiple cache misses for each particle that severely hits the performance.
Motivated from \citet{2017CoPhC.210..145V}, we
create what we call the {\it intermediate arrays}.
On top of the existing grid data, it creates the innermost dimension which gather/record information from/for neighboring grid points along the $x_2-$direction, as shown in Figure~\ref{fig::interm}. Effectively, the size of this array is three times the original data in 2D, and nine times in 3D, while this array is not needed in 1D as the grid data for interpolation is already contiguous in memory.
When interpolating for / depositing from the particle in Figure~\ref{fig::par_3d}, the program first finds the data from the three gird points sharing the same $x_1-$index, then continuously accesses the data of the other six grid points from the intermediate array. Algorithm~\ref{alg::mapping} provides the mapping relation between the intermediate arrays and the original grids. The mapping process itself is slow due to constant non-contiguous data access, and costs a substantial overhead.
On the other hand, such cost is usually negligible when there is a large ($\gtrsim$ a few tens) number of particles per cell that is typical in (MHD)-PIC applications, giving way to the major benefit through enhancing the vectorization efficiency of the particle module.

Here we describe our vectorization strategy for the numerical stages in Section~\ref{sec::num_mhd-pic}. One vectorized loop can simultaneously processes \texttt{SimdWidth} number of data
(\texttt{SimdWidth}=8 in the latest processors), under one single instruction.
We package the simulation particles in groups of \texttt{SimdWidth}
and employ a hybrid vectorization strategy for particles.
In Stage~1.2~,~2.2~and~2.4, we loop over the particles in each group, while aim to vectorize over the intermediate arrays for interpolation and deposits.
Such design avoids the data conflict and achieves the best vectorization efficiency. The handling of boundary communications (Stage~1.5~\&~2.7) is not vectorized, as it is conditional on the location of simulation particles.

\begin{algorithm}[!ht] 
	\caption{Mapping relations for the intermediate array}
 	\label{alg::mapping}
\DontPrintSemicolon
  \SetKwFor{For}{for}{}{}
  \SetKwData{Npar}{Npar}
  \SetKwData{Loc}{Pos}
  \SetKwData{Mom}{Mom}
  \SetKwData{Idx}{Crd}
  \SetKwData{Dt}{Dt}
  \SetKwData{Wgt}{Weight}
  \SetKwData{MHD}{Bkgnd}
  \SetKwData{N}{[n]}
  \SetKwData{U}{U}
  \SetKwData{B}{B}
  \SetKwData{StartPar}{Start}
  \SetKwData{EndPar}{End}
  \SetKwData{Change}{DeltaMom}
  \SetKwData{Feedback}{Fdbk}
  \SetKwData{IntermFdbk}{IntermFdbk}
  \SetKwData{IntermPrim}{IntermBkgnd}
  \SetKwData{SimdWidth}{simd\_width}
  \SetKwData{NpcCube}{ParCloud}
  \KwData{\\
  \Feedback  \tcp*{CRs' feedback on MHD grids}
  \MHD       \tcp*{gas velocity or magnetic field on MHD grids}
  \IntermFdbk \tcp*{Intermediate array of \Feedback}

  \IntermPrim \tcp*{Intermediate array of \MHD}

  \NpcCube \tcp*{Particle could size, =27 in 3D simulations, =9 in 2D and =3 in 1D}
  }
  \KwResult{Copying \MHD to \IntermPrim, and summing up \IntermFdbk to \Feedback}
  \For{k = 0 \KwTo $N_z$}{
	\For{j = 0 \KwTo $N_y$}{
		\For{i = 0 \KwTo $N_x$}{
			\For{d = 0 \KwTo \NpcCube / 3}{
				\IntermPrim[k][j][i][d] = \MHD[k + d / 3 - 1][j + d \%3 - 1][i]\;
			}
		}
    }  
  }

  \tcp*{\Feedback has been initialized with 0}
  \For{k = 0 \KwTo $N_z$}{
	\For{j = 0 \KwTo $N_y$}{
		\For{i = 0 \KwTo $N_x$}{
			\For{d = 0 \KwTo \NpcCube / 3}{
				\Feedback[k + d / 3 - 1][j + d \%3 - 1][i] += \IntermFdbk[k][j][i][d]\;
			}
		}
    }  
  }
\end{algorithm}

The intermediate arrays enhance the cache hit rate when interpolating or depositing for one single particle, but cache misses are likely to arise when the program proceeds to the next particle, which in the general case can be located at a very different position that must access grid data far from those used by the previous particle.
A common optimization strategy is to sort the particles according to their spatial locations (e.g. \citealp{2019CoPhC.244..246B,2021arXiv210213133B}).
We employ the sort algorithm offered by the standard template library in \texttt{C++} \footnote{\citet{2019CoPhC.244..246B} provided a sorting algorithm that is the most efficient for simulations where particle velocities are relatively low, although it is not necessarily advantageous in our case as MHD-PIC applications typically targets at (trans-)relativistic CR particles with relative high velocities.}.
Note that particle sorting is relatively computationally expensive, we thus only do particle sorting over the interval of many integration timesteps. The exact interval to optimize the performance is problem dependent. For instance, with a largely uniform particle distribution using $10^3$ particle per cell (ppc), we find that sorting the particles every $\sim 200$ timesteps, where $\sim$10\% simulation particles have crossed into the neighbour MeshBlocks, is nearly optimal for the electrons’ and positrons’ gyration test (Section~\ref{sec::rel_vel}).

\begin{figure}
	\includegraphics[width=\columnwidth]{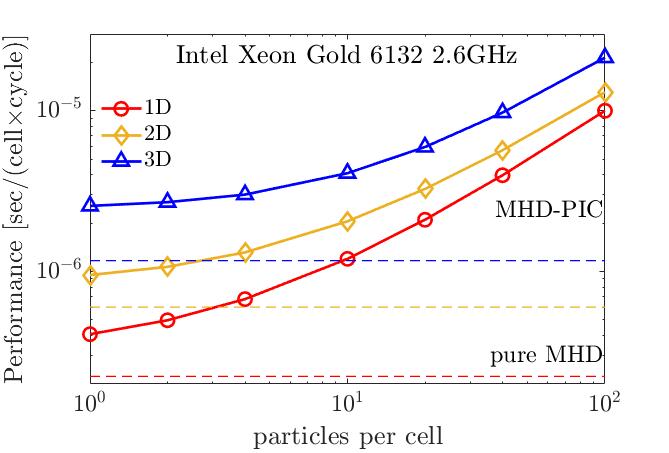}
    \caption{The performance for MHD-PIC on Intel Xeon Gold 6132 2.6GHz, with the test in Section~\ref{sec::test_bell}. The total time cost for integrating one cell relies on the number of particle per cell, as the sum of pure MHD cost and particle cost. The precision is in double and the simulation dimensions are referred to colors.}
    \label{fig::performance}
\end{figure}

We measure our code performance using the Bell instability test (Section~\ref{sec::test_bell}) in double precision, using the Intel compiler and all cores in one computer node with 2 Intel Xeon Gold 6132 CPUs (totally 28 cores). We vary the number of particle per cell and conduct the test in 1,2 and 3 dimensions. Figure~\ref{fig::performance} shows the
computational cost in terms of time (second) per cell updates per timestep in solid lines. For comparison, the cost of identical but pure MHD simulations are shown in dashed lines. The dependence of the cost on number of particles per cell is more flat for small ppc$\lesssim10$, reflecting the constant overhead associated with the intermediate array, which itself can be $\lesssim 2$ times the cost of the MHD integrator itself.
On average, the cost of updating one MHD cell (excluding the overhead from the intermediate array) corresponds to updating $\sim2$ particles in 1D, $\sim4$ particles in 2D, and $\sim6$ particles in 3D.

\subsection{Parallelization and load balancing}\label{sec::load}
\begin{figure}
	\includegraphics[width=\columnwidth]{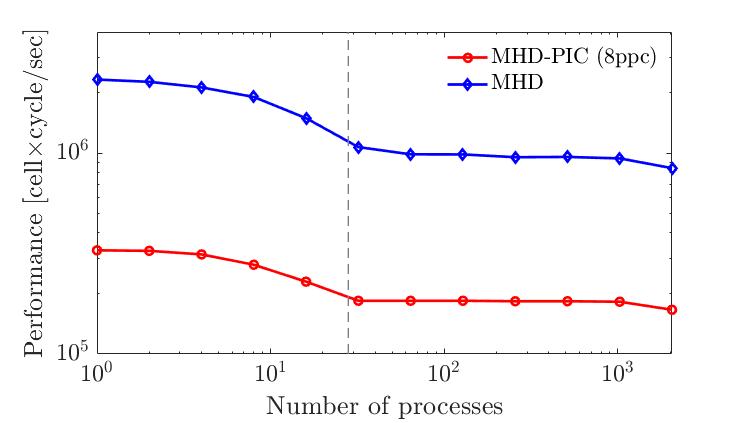}
    \caption{Results of the weak scaling test on our local computer cluster with 80 computing nodes with 2 Intel Xeon Gold 6132 CPUs (28 cores) per node. The vertical axis indicates the cell number per core that the program integrates per second. The red markers are the results using the test in Section~\ref{sec::rel_vel} with 8 ppc and the blue ones represent the results from a simple MHD advection problem. The vertical dashed line restricts the single node performance on its left.}
    \label{fig::weakscale}
\end{figure}

\texttt{Athena++} employs a hybrid parallelization strategy, compatible with both the Message Passing Interface (MPI) and the OpenMP standard \citep{2020ApJS..249....4S}. Parallel threads/processes separately integrate the MeshBlocks. While data communication between MeshBlocks can be intensive. \texttt{Athena++} employs a dynamical scheduling system called the ``TaskList", which is coupled with non-blocking MPI operations to hide the communication latency (see more details in Section 2.4 from \citealp{2020ApJS..249....4S}). Following the same design, we embed non-blocking MPI communication for the simulation particles in the TaskList. To test the parallelization efficiency, we employ the test problem in Section~\ref{sec::rel_vel} and show the weak scaling test result in Figure~\ref{fig::weakscale}. In the test, we choose a diagonal background field so that particles can enter all neighboring MeshBlocks during gyration. Each process handles a total of $64^3$ cells divided into eight MeshBlocks and each cell contains eight particles. We compile the code with the \texttt{Intel} compiler and run the program on a computer cluster with 2 Intel Xeon Gold 6132 CPUs per node (totally 28 cores per node) using up to 74 nodes. A pure MHD advection problem is run in the same grid design for comparison. The computation efficiency, for both MHD and MHD-PIC, decreases with increasing number of processes up within a single node, while remains largely flat afterwards. Overall, the parallelization efficiency of the MHD-PIC module among nodes is largely identical to that of the MHD part in the original \texttt{Athena++} code.

\texttt{Athena++} offers the load balancing attribute which dynamically re-distributes the workload as evenly as possible to all processes (see more details in Section~2.1.6 from \citealp{2020ApJS..249....4S}). Before the re-distribution, the program first predicts the workload of each MeshBlock according to the CPU time in the last step. The MHD-PIC workload consists of the MHD integration, the overhead from the intermediate arrays and the particle integration. The first two parts scale with the number of cells, which are similar among MeshBlocks at different levels. The cost from particle integration is largely proportional to the number of particles residing in the MeshBlock. In our implementation, we employ a rough estimate based on Figure~\ref{fig::performance}, where the computational cost per cell for the first two parts corresponds to the integration of about $\sim$3 particles in 1D, $\sim$7 in 2D, and $\sim$12 in 3D\footnote{Note that the computational cost ratio between one cell and one particle depends on the compiler. Figure~\ref{fig::performance} is based on the Intel compiler on Intel processors. The GNU compiler generally yields a smaller ratio on the same processors. The users can adjust these ratios to further optimize load balancing in their systems.}, to estimate the MHD-PIC workload of individual MeshBlocks, as well as for refined/de-refined MeshBlocks for AMR. An example for the load balancing is provided in the shock acceleration simulations (Section~\ref{sec::test_shock}). We also note that load balancing can be applied regardless of whether there is mesh refinement or not, because individual MeshBlocks may possess highly different number of particles.

\subsection{Mesh refinement for particles}
\label{sec::refinement}

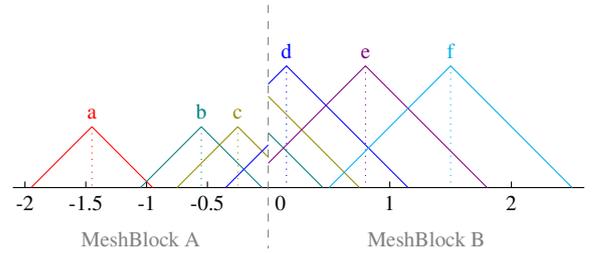
\begin{figure}
	\centering
	\begin{tikzpicture}[scale=1.6]
		\draw (-2.1,0) -- (2.6,0);
		\draw (-2.0,0.05) -- (-2.0,0) node[below] {-2};
		\draw (-1.5,0.05) -- (-1.5,0) node[below] {-1.5};
		\draw (-1.0,0.05) -- (-1.0,0) node[below] {-1};
		\draw (-0.5,0.05) -- (-0.5,0) node[below] {-0.5};
		\draw (0.1,0) node[below] {0};
		\draw[gray] (-1.05,-0.3) node[below] {MeshBlock A};
		\draw[gray] (1.3,-0.3) node[below] {MeshBlock B};
		\draw[dashed,gray] (0,1.5)--(0,-0.5);
		\draw (1.0,0.05) -- (1.0,0) node[below] {1};
		\draw (2.0,0.05) -- (2.0,0) node[below] {2};
		\draw[color=red] (-1.95,0)--(-1.45,0.5) node[above]{a};
		\draw[color=red] (-1.45,0.5)--(-0.95,0);
		\draw[dotted,red] (-1.45,0.5)--(-1.45,0);
		\draw[color=teal] (-1.05,0)--(-0.55,0.5) node[above]{b};
		\draw[color=teal] (-0.55,0.5)--(-0.05,0);
		\draw[color=teal] (0,0.45)--(0.45,0);
		\draw[dotted,teal] (-0.55,0.5)--(-0.55,0);
		\draw[color=olive] (-0.75,0)--(-0.25,0.5) node[above]{c};
		\draw[color=olive] (-0.25,0.5)--(0,0.25);
		\draw[color=olive] (0,0.75)--(0.75,0);
		\draw[dotted,olive] (-0.25,0.5)--(-0.25,0);
		\draw[color=blue] (0,0.85)--(0.15,1.0) node[above]{d};
		\draw[color=blue] (0.15,1)--(1.15,0);
		\draw[color=blue] (0,0.35)--(-0.35,0);
		\draw[dotted,blue] (0.15,1)--(0.15,0);
		\draw[color=violet] (0,0.2)--(0.8,1.0) node[above]{e};
		\draw[color=violet] (0.8,1)--(1.8,0);
		\draw[dotted,violet] (0.8,1)--(0.8,0);
		\draw[color=cyan] (0.5,0)--(1.5,1.0) node[above]{f};
		\draw[color=cyan] (1.5,1)--(2.5,0);
		\draw[dotted,cyan] (1.5,1)--(1.5,0);
	\end{tikzpicture}
	\caption{A sketch for particles' deposit weight under refinement. The $x$-axis refers to the spatial location, where each segment refers to one cell. Two neighboring MeshBlocks are at two refinement levels. Cells in the finer MeshBlock `A' 
        have negative coordinates with the half cell size compared with that of the coarser MeshBlock `B'. We consider six particles, whose TSC weight functions are marked in different colors, 
        where the weights for individual cells are proportional to the area that the solid lines intersect with each cell. The location of the six particles are -1.45, -0.55, -0.25, 0.15, 0.8 and 1.5 respectively. Particles `a' and `f' are not affected by refinement, thus the total deposit weights sum to 1. To ensure smoothness of weighting functions as particles cross fine-coarse MeshBlock boundaries, the weight function in MeshBlock `A' is normalized by the full triangular area of particle `a' and that in MeshBlock `B' is normalized by the full triangular area of particle `f'.}
	\label{fig::refinement}
\end{figure}
Our MHD-PIC code supports both the static and adaptive mesh refinement.\footnote{We acknowledge the contribution from C.-C. Yang for sharing an early version of the particle module, from which we largely inherited the treatment of mesh refinement, and greatly benefit from the handling of data communications.} The different resolution across refinement boundaries requires further consideration on the interpolation and deposit schemes. In \texttt{Athena++}, neighboring MeshBlocks can differ by at most one refinement level, whose grid size differ by a factor of two. Therefore, it suffices to consider the interpolation scheme between two fine and coarse MeshBlocks.

In our numerical treatment, the interpolation scheme is identical to the case without refinement.
This is because each MeshBlock is surrounded by at least two layers of ghost cells, whose values are filled through restriction/prolongation when they are located at fine/coarse boundaries. This ensures interpolation to proceed in the usual way for particles near MeshBlock boundaries.

The deposit around the interfaces between different refinement levels is less straightforward. The bottom line here is that for
a uniform particle distribution, the particle deposit should ensure homogeneity across fine/coarse boundaries. As a result, for particles near the refinement boundaries, one must explicitly consider different resolutions across the boundary. Our solution here is to set the weights in each cell according to the TSC weight function {\it at the resolution of that cell}.
Figure~\ref{fig::refinement} presents a sketch of the deposit weight function in 1D, including six mock particles near the boundary between two MeshBlocks at two refinement levels.
Refinement does not affect the particle `a' in the fine level and the particle `f' in the coarse level, where the weights are proportional to the area under colored solid lines in each cell, normalized by the total area of the triangular cloud. These weights sum up to 1.
The particle `b', `c' and `d' give their feedback to cells in both MeshBlocks with different refinement levels. Let us take particle `c' as an example. It deposits to the two cells in Meshblock A as usual. We then imagine this particle `c' to reside in the ghost zones of Meshblock B at coarse level to compute the TSC weights to the remaining cell in Meshblock B. Note that under this treatment, the total weights do not sum up to 1 (exceed 1 in this case).
Another special case is particle `e', which only affects two cells in Meshblock B and the total weights is less than 1. This is because it is too far from Meshblock A should it reside in the fine level. 

Obviously, our treatment gives up the total momentum (energy) conservation from individual particles where the feedback to gas may be greater or less than the change in particle momentum (energy). On the other hand, it is more important to ensuring smoothness across refinement boundaries to avoid numerical artifacts.
We also verify our implementation with two test problems in Section~\ref{sec::rel_vel}~\&~\ref{sec::test_shock}, both of which meet our expectations.

\subsection{The $\delta f$ method}\label{ssec:deltaf}
In Equation~\ref{equ::mhd_momentum}~\&~\ref{equ::mhd_energy}, the full CR distribution function is represented by a finite number of simulation (super-)particles. The CR feedback, given by certain integral over the CR distribution function, is approximated by deposits from finite number of particles. For instance, the CR number density at location $\boldsymbol{x}$ is given by,
\begin{equation*}
    \int f_\alpha \left(t, \boldsymbol{x}, \boldsymbol{p}\right) \dd^3 \boldsymbol{p} \approx  \sum_{k} S\left(\boldsymbol{x} - \boldsymbol{x}_{\alpha, k} \right),
\end{equation*}
where $S\left(\boldsymbol{x} - \boldsymbol{x}_k \right)$ is the TSC shape function of particle $k$, also seen as the solid lines in Figure~\ref{fig::refinement}, and $\alpha$ refers to the particle species. The deposit value is statistical, following a Poisson distribution whose error scales as the inverse square root of the number of particles per cell.

In certain applications, when the background distribution $f_0$ is known and the full distribution $f\left(t, \boldsymbol{x}, \boldsymbol{p} \right)$ is not expected to strongly deviate from $f_0$, one can dramatically reduce the Poisson noise by employing the $\delta f$ method \citep[e.g.][]{1993PhFlB...5...77P, 1994PhPl....1..863H, 1995JCoPh.119..283D, 2014JCoPh.259..154K, 2019ApJ...876...60B}. The CR distribution is divided into two parts, a background distribution $f_0$
and deviations from it $\delta f \equiv f - f_0$.
The CR source terms in Equation~\ref{equ::mhd_momentum}~\&~\ref{equ::mhd_energy} thus break up into two parts,
\begin{align}
	\frac{\rho_{CR}}{c} c\boldsymbol{E} + \frac{\boldsymbol{j}_{CR}}{c} \times \boldsymbol{B} =& \frac{\left(\rho_{CR,0} + \rho_{\delta}\right)}{c} c\boldsymbol{E} + \frac{\left(\boldsymbol{j}_{CR,0} + \boldsymbol{j}_{\delta}\right)}{c} \times \boldsymbol{B}, \notag \\
	\frac{\boldsymbol{j}_{CR}}{c} \cdot \left(c\boldsymbol{E}\right) =& \frac{\left(\boldsymbol{j}_{CR,0} + \boldsymbol{j}_{\delta}\right)}{c} \cdot \left(c\boldsymbol{E}\right), \label{equ::delta_f_source} 
\end{align}
where $\rho_{CR,0}$ and $\boldsymbol{j}_{CR,0}$ denote the charge density and current density from the $f_0$ respectively, and can be calculated analytically. 
While the CRs are Lagrangian tracers of the full distribution function, they are employed to evaluate
the $\delta f$ part, $\rho_\delta$ and $\boldsymbol{j}_\delta$
\begin{align}
	\frac{\rho_{\delta}}{c} \approx & \sum_{\alpha, k} \left(\frac{q}{mc}\right)_\alpha m_\alpha w_k S\left(\boldsymbol{x} - \boldsymbol{x}_k \right), \notag \\
	\frac{\boldsymbol{j}_{\delta}}{c} \approx & \sum_{\alpha, k} \left(\frac{q}{mc}\right)_\alpha m_\alpha w_k \boldsymbol{v}_k S\left(\boldsymbol{x} - \boldsymbol{x}_k \right),
\end{align}
with the weight $w_k$ 
for simulation particle $k$ given by
\begin{equation*}
	w_k = \frac{\delta f\left(t, \boldsymbol{x}_k\left(t\right), \boldsymbol{p}_k\left(t\right)\right)}{f\left(t, \boldsymbol{x}_k\left(t\right), \boldsymbol{p}_k\left(t\right)\right)} = 1- \frac{f_0\left(t, \boldsymbol{x}_k\left(t\right), \boldsymbol{p}_k\left(t\right)\right)}{f\left(0, \boldsymbol{x}_k\left(0\right), \boldsymbol{p}_k\left(0\right)\right)},
\end{equation*}
where
the Liouville theorem is applied that the full $f$ is conserved
along 
particle trajectories
in phase space, $f\left(t, \boldsymbol{x}_k\left(t\right), \boldsymbol{p}_k\left(t\right)\right)=f\left(0, \boldsymbol{x}_k\left(0\right), \boldsymbol{p}_k\left(0\right)\right)$. 
The CR deposits in the
second stage during particle integration can be treated in a similar fashion. 
For $f$ approaching $f_0$,
namely $w_k \ll 1$, the deposit
directly from the particles is substantially suppressed compared to the treatment in the standard ``full-$f$" method, where $w_k=1$, thus dramatically reduces the noise. We also that the $\delta f$ method no longer conserves energy and momentum to machine precision due to the variable weights, but generally such errors are reduced together with numerical noise. On the other hand, the advantage of $\delta f$ vanishes when $f$ strongly deviates with $f_0$.

The $\delta f$ method in MHD-PIC is important for studying the microphysics of CR transport, where the background CR distribution is nearly isotropic. We will provide a set of test problems in Section \ref{sec::test_gyro-resonance} in this area.

\section{Benchmark tests}
In this section, we present a series of benchmark test problems to examine the accuracy and performance of various ingredients of our MHD-PIC implementation.

\subsection{Gyration}
\label{sec::gyro}
\begin{figure}
	\includegraphics[width=\columnwidth]{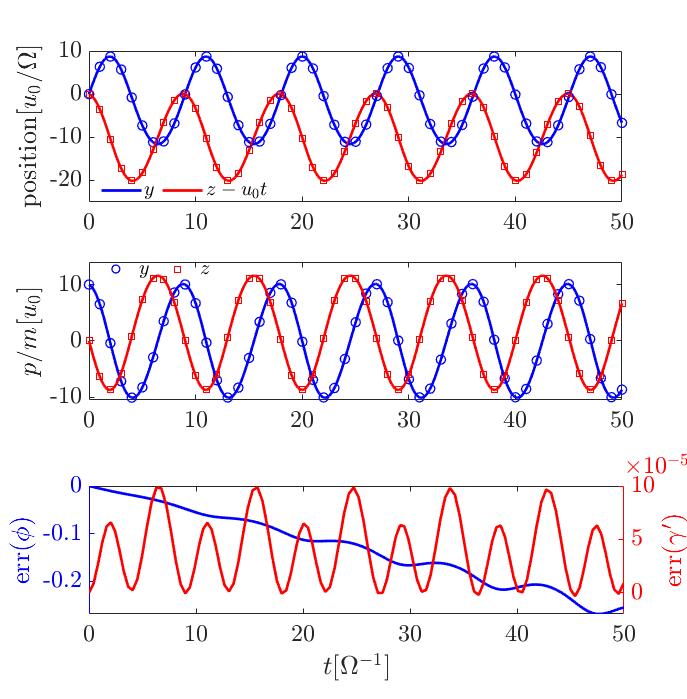}
    \caption{The position (top) and momentum (middle) of one CR particle in a uniform electromagnetic field as a function of time. Markers are from simulation data and solid lines represent the analytic solution from Equations~\ref{equ::gyro_analytical_configure}~\&~\ref{equ::gyro_analytical_momentum}. Blue and red colors represent $y$ and $z$ components. The relative errors are shown in the bottom panel. The gyro phase error corresponds to the blue line (left $y$-axis), while the particle energy error in the gas co-moving frame are shown in the red line (right $y$-axis).}
    \label{fig::gyration}
\end{figure}
To test the accuracy of the Boris pusher,
we set up a uniform background gas and inject a CR particle without feedback. The gas moves in a velocity $u_0 \hat{z}$ with a background field $B_g\hat{x}$. The CR has an initial momentum $\left(\boldsymbol{p}/m\right)\left(t=0\right)=v_0\hat{y}$. In non-relativistic limit, $u_0 \ll \mathbb{C}$ and $v_0 \ll \mathbb{C}$, the CR makes gyro motion with frequency $\Omega = qB_g/\left(mc\right)$ plus a drift velocity $u_0 \hat{z}$. To be general, we simulate a trans-relativistic case where $\mathbb{C}=v_0=10u_0$. In the co-moving frame where the gas is static, the CR particle undergoes orbital motion with gyro frequency $\Omega \left(1-u_0^2/\mathbb{C}^2\right)/\sqrt{1+v_0^2/\mathbb{C}^2}$. After the Lorentz transform, we boost CR's position and momentum back to the simulation frame, and the result is
\begin{align}
	\boldsymbol{x} = \left(
	\begin{aligned}
	& 0 \\
	& \frac{\sqrt{v_0^2+u_0^2} \sin\phi - \sqrt{1 + v_0^2/\mathbb{C}^2} u_0}{\left(1-u_0^2/\mathbb{C}^2\right)\Omega} \\
	& u_0 t  + \frac{\sqrt{v_0^2+u_0^2}}{\sqrt{1-u_0^2/\mathbb{C}^2}\Omega} \cos\phi - \frac{v_0}{\Omega}
	\end{aligned}
	\right), \label{equ::gyro_analytical_configure} 
\end{align}
\begin{align}
	& \frac{\boldsymbol{p}}{m} = \left(
	\begin{aligned}
	& 0\\
	& \sqrt{\frac{v_0^2+u_0^2}{1-u_0^2/\mathbb{C}^2}} \cos\phi \\
	& u_0 \frac{\sqrt{1 + v_0^2 /\mathbb{C}^2}}{1-u_0^2/\mathbb{C}^2} -\frac{\sqrt{v_0^2+u_0^2}}{1-u_0^2/\mathbb{C}^2} \sin\phi
	\end{aligned}
	\right), \label{equ::gyro_analytical_momentum} \\
	&\phi\left(t,z\right) \equiv \Omega \sqrt{\frac{1-u_0^2/\mathbb{C}^2}{1+v_0^2/\mathbb{C}^2}} \left(t - \frac{u_0 z}{\mathbb{C}^2}\right) + \arccos \sqrt{\frac{1-u_0^2/\mathbb{C}^2}{1+u_0^2 / v_0^2}}. 
    \label{equ::gyro_analytical_phi}
\end{align}
The CR motion is still periodic but not sinusoidal in the $y-z$ plane, as gyro phase $\phi$ depends on location as well.

In the simulation, we
adopt the computational unit of $u_0=B_g=q/\left(mc\right)=\Omega=1$. The simulated box is cubic whose size is $500u_0 / \Omega$ in each direction with 32 cells. We restrict that in each integration step the CR cannot rotate by more than $\theta_\text{max}=0.3$ radian.
To compare simulation results with the analytical solution, we solve for $z$ as a function of $t$, and plug in $z(t)$ to Equation~\ref{equ::gyro_analytical_phi} to obtain the full analytical solution. The numerical results, integrated for $50 \Omega^{-1}$, very well converge to the analytical solutions (Equation~\ref{equ::gyro_analytical_configure}~\&~\ref{equ::gyro_analytical_momentum}) as shown in Figure~\ref{fig::gyration}, where we have subtracted the gas motion $u_0 t$ from the CR particle displacement in the $z-$direction. CR particle energy $\gamma'$ in the gas co-moving frame is largely conserved despite of the small-amplitude oscillations. We also see that the truncation error mainly exhibits as a phase error. All these properties conform to the standard Boris pusher.

\subsection{The Bell instability}
\label{sec::test_bell}

\begin{figure}
	\includegraphics[width=\columnwidth]{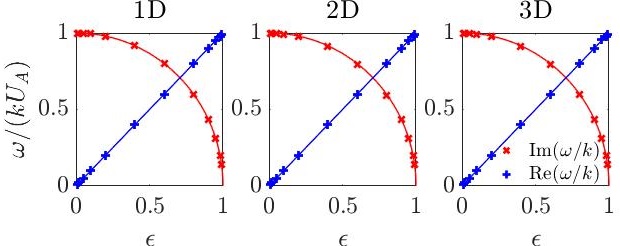}
    \caption{The linear dispersion relation of the Bell instability in 1D, 2D, and 3D. Markers represent simulation data and lines correspond to analytical results from Equation~\ref{equ::bell_unstable}. Red and blue correspond to imaginary parts (wave growth rates) and real parts (phase speeds) of the wave frequency respectively, which only depend on the ratio of the Alfv\'en speed and the CR streaming speed. $\epsilon \equiv U_\text{A} / v_\text{CR}$.}
    \label{fig::bell}
\end{figure}

Next, we turn on CR feedback and conduct
the Bell instability \citep{2004MNRAS.353..550B} test.
It occurs in the presence of a strong CR current $\boldsymbol{j}_\text{CR}$ streaming through the background magnetic field $\boldsymbol{B}_g$, and trigger the growth of right-polarized circularly polarized Alfv\'en waves. It is a non-resonant instability, meaning that
the unstable wavelengths are much shorter than the CR gyro-radii.
Its dispersion relation reads \citep[here we adopt the form given in ][]{2015ApJ...809...55B}:
\begin{align}
	\left(\omega \pm \epsilon k_0 U_\text{A}\right)^2 = \left(k\pm k_0\right)^2 U_\text{A}^2 - \left(1-\epsilon^2\right) k_0^2 U_\text{A}^2, \label{equ::bell_dispersion} \\
	k_0 \equiv \frac{j_\text{CR}}{2B_g c}, \quad \epsilon \equiv \frac{U_\text{A}}{v_\text{CR}}. \notag
\end{align}
where $\epsilon$ corresponds to the dimensionless streaming speed. 
The instability is triggered when the CR streaming velocity $v_\text{CR}$ exceeds the Alfv\'en velocity $U_A$, with the most unstable wavenumber at $k=k_0$.

In our simulations, we initialize an eigenmode of the most unstable mode, $k=k_0$ \citep[given in ][]{2015ApJ...809...55B}. The direction of the wave lies in the $x$-axis in 1D simulations and are along the diagonal in 2D or 3D. The gas velocity and magnetic fluctuations
follow the relation
\begin{equation}
	\delta \boldsymbol{u} = \frac{\delta \boldsymbol{B}}{\sqrt{\rho}} \exp \bigg[i \left(-\arcsin \epsilon - \frac{\pi}{2}\right)\bigg]. \label{equ::bell_unstable}
\end{equation}
All other MHD variables are uniform.
In the simulations, we uniformly assign two particles per cell.
The computational units are $\rho=B_g=U_\text{A}=2 \pi / k_0 = 1$. The simulation box size and resolution $\left(L_x / N_x, L_y / N_y, L_y / N_y \right)$ are $\left(1 / 32\right)$ in 1D, $\left(\sqrt{5} / 64, \sqrt{1.25} / 32\right)$ in 2D, and $\left(\sqrt{21} / 128, \sqrt{5.25} / 64, \sqrt{1.3125} / 32 \right)$ in 3D, where the wavelength is fixed to be 1. 
To suppress the resonant streaming instability (see Section~\ref{sec::test_gyro-resonance}),
we set the gyro frequency to be lower than the Alfv\'en wave frequency by six order of magnitude, and the numerical speed of light to be much greater than the CR streaming speed, $\Omega\equiv q B_g / mc = 10^{-6} k_0 U_\text{A}$ and $\mathbb{C}= 10 ^3 v_\text{CR}$.

We aim to test the dispersion relation at the most unstable wavelength, where
Equation~\ref{equ::bell_dispersion} leads to:
\begin{equation*}
	\omega = k U_\text{A} \left(\epsilon + i\sqrt{1-\epsilon^2}\right).
\end{equation*}
The real part controls the phase velocity and the imaginary part corresponds to the wave growth rate. We measure the change in wave phase over constant time intervals $\pi / \left(k_0 U_A\right)$ by spatially fitting $\delta u_y$ with a sine function, and calculate the real part of $\omega$. The instability growth rate is measured through fitting the growth of volume-averaged $|\boldsymbol{\delta u}|$ over time by an exponential function. 
The results are shown in Figure ~\ref{fig::bell}. 
A smaller $\varepsilon$, namely a faster CR streaming speed,
leads to higher growth rate and slower propagation.
Clearly, our simulation results well agree with the analytical dispersion relation at all dimensions, even with modest resolution.

\subsection{Oscillation of gas with electrons and positrons}
\label{sec::rel_vel}

\begin{figure}
	\includegraphics[width=\columnwidth]{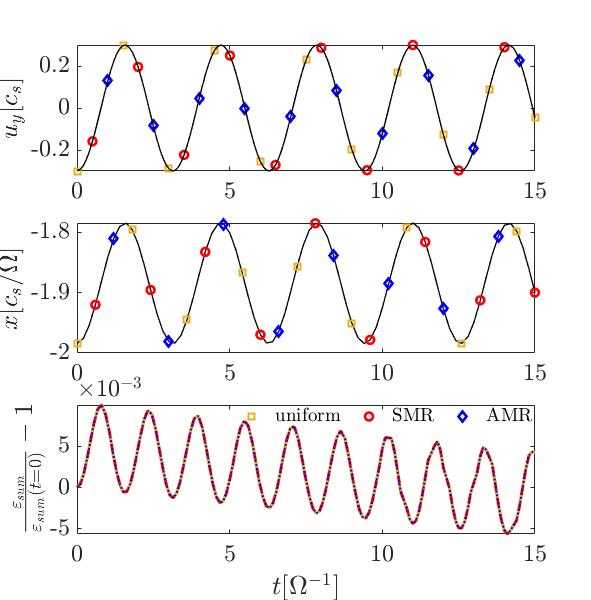}
    \caption{Test result for gas oscillation with electrons and positrons. The top panel is the time evolution of the volume-averaged $u_y$, the gas velocity in $\hat{y}$. The middle panel shows one particle trajectory in the $x-$direction. The bottom panel shows the numerical error of gas kinetic energy density $\varepsilon_\text{sum}$. In the top two panels, markers are simulation data and the solid lines represent the analytic result (Equation~\ref{equ::rel_vel}). Colors correspond to the grid setups, corresponding to whether local mesh refinement is enabled (to test homogeneity in particle deposit across refinement boundaries), where in the bottom panel lines with different grid setups overlap with each other.}
    \label{fig::rel_vel}
\end{figure}

Here we present
another test to verify CR backreaction with multiple particle species, where the MHD gas oscillates together with the gyration of hot electrons' and positrons'. In a uniform magnetic field $\boldsymbol{B}_g$, hot electrons and positrons, denoted by $-$ and $+$ respectively, make gyro motions and exert feedback to the thermal gas. In the non-relativistic limit, $\gamma\approx 1$, the gyro frequency is always the same constant for both electrons and positrons, $\Omega=B_g q / \left(m_ec\right)$. For simplicity, we require the mass density for electrons and positrons are uniform and identical, denoted as $m_e n_0=\rho_-=\rho_+$. The dynamical equations in the center-of-mass frame of the system are
\begin{align}
	&\frac{\dd \boldsymbol{v}_+}{\dd t} = \frac{q}{m_ec} \left(\boldsymbol{v}_+ - \boldsymbol{u}\right)\times \boldsymbol{B}_g, \notag \\
	&\frac{\dd \boldsymbol{v}_-}{\dd t} = -\frac{q}{m_ec} \left(\boldsymbol{v}_- - \boldsymbol{u}\right)\times \boldsymbol{B}_g, \notag \\
	&m_e n_0 \left(\boldsymbol{v}_+ + \boldsymbol{v}_-\right) + \rho \boldsymbol{u} = 0. \label{equ::rel_vel}
\end{align}
As everything is uniform, $\boldsymbol{B}_g$ and $\rho$ do not vary with space and time. We obtain after some algebra
\begin{equation}
	\partial_t^2 \boldsymbol{u} = -\Omega^2 \left(1 + 2m_e n_0 / \rho\right) \boldsymbol{u}.
    \label{equ::rel_vel_gas_osc}
\end{equation}
Therefore, the characteristic oscillation frequency of the system is given by
$\Omega \sqrt{1 + 2m_e n_0 / \rho}$.

To realize this oscillation,  we uniformly fill a 3D periodic box with electrons and positrons, with 64 ppc per species, whose $m_e n_0 / \rho = 1.5$ (not physically realistic, but it suffices for test purposes). We use an isothermal equation of state with sound speed $c_s = 1$ in code units, together with $\rho=q / \left(m_ec\right) = B_g =\Omega = 1$. Both electrons and positrons share the same initial velocity $\boldsymbol{v}_+ = \boldsymbol{v}_-=0.1 c_s \hat{y}$, and the initial magnetic field is along the $z$-direction. In the center-of-mass frame,
the gas moves
with an opposite $y-$ velocity $\boldsymbol{u}=-0.3 c_s\hat{y}$.
We fix $\mathbb{C} = 10^3 c_s$, to ensure the system is well in the non-relativistic limit. We also restrict the integration time step $\Delta t \Omega \sim 10^{-1}$.\footnote{While $\Delta t \Omega \sim 0.3$ generally yields sufficiently accurate CR gyration orbit, we find for this test that a smaller $\Delta t$ is needed for stability, and this is likely due to the (unrealistically) strong CR mass loading.}

We compare the gas oscillation in the simulation with the analytical solution in Figure~\ref{fig::rel_vel}. The MHD gas should uniformly oscillate in the $y-$ direction, with the frequency $\Omega \sqrt{1 + 2m_e n_0 / \rho}$. The volume-averaged gas velocity $u_y(t)$ in the simulation coincides with the expectation from Equation~\ref{equ::rel_vel_gas_osc}. We examines the error in kinetic energy, which should be conserved
\begin{equation*}
    \varepsilon_\text{sum}\equiv \frac{\rho u_y^2}{2} + \frac{m_e n_0}{2} \left(v_{+}^2+v_{-}^2\right).
\end{equation*}
The relative error is small, primarily in the form of a phase error fluctuating around zero. We also note that while in principle the gas should only have $y-$ velocity, small $x-$velocity is also developed as the main source of error (thus the decreasing trend in $\varepsilon_{\rm sum}$).

To test CR backreaction in the presence of mesh refinement, we add one refinement level (two levels in total), where the size of coarser cells, the number of particle per cell in the coarser level and the $\Delta t$ are all identical to those in the uniform grid run. We refine one eighth of the total region in the static mesh refinement (SMR) run. In the adaptive mesh refinement (AMR) run, the program randomly sets the refinement criteria, with 10\% possibility to refine and 60\% possibility to derefine into a coarser level after each computation step. On average, about 82\% physical area is in the finer level. The SMR run and the AMR run both reproduce the gas oscillation as well as individual particle trajectories, as seen in Figure~\ref{fig::rel_vel}. We see that the results are identical to that in a uniform grid, even sharing the same numerical error.
The results confirm that feedback from a uniform spatial distribution of particles is well preserved in the presence of mesh refinement.

\subsection{Non-relativistic shock acceleration}
\label{sec::test_shock}

\begin{figure*}
	\includegraphics[width=\textwidth]{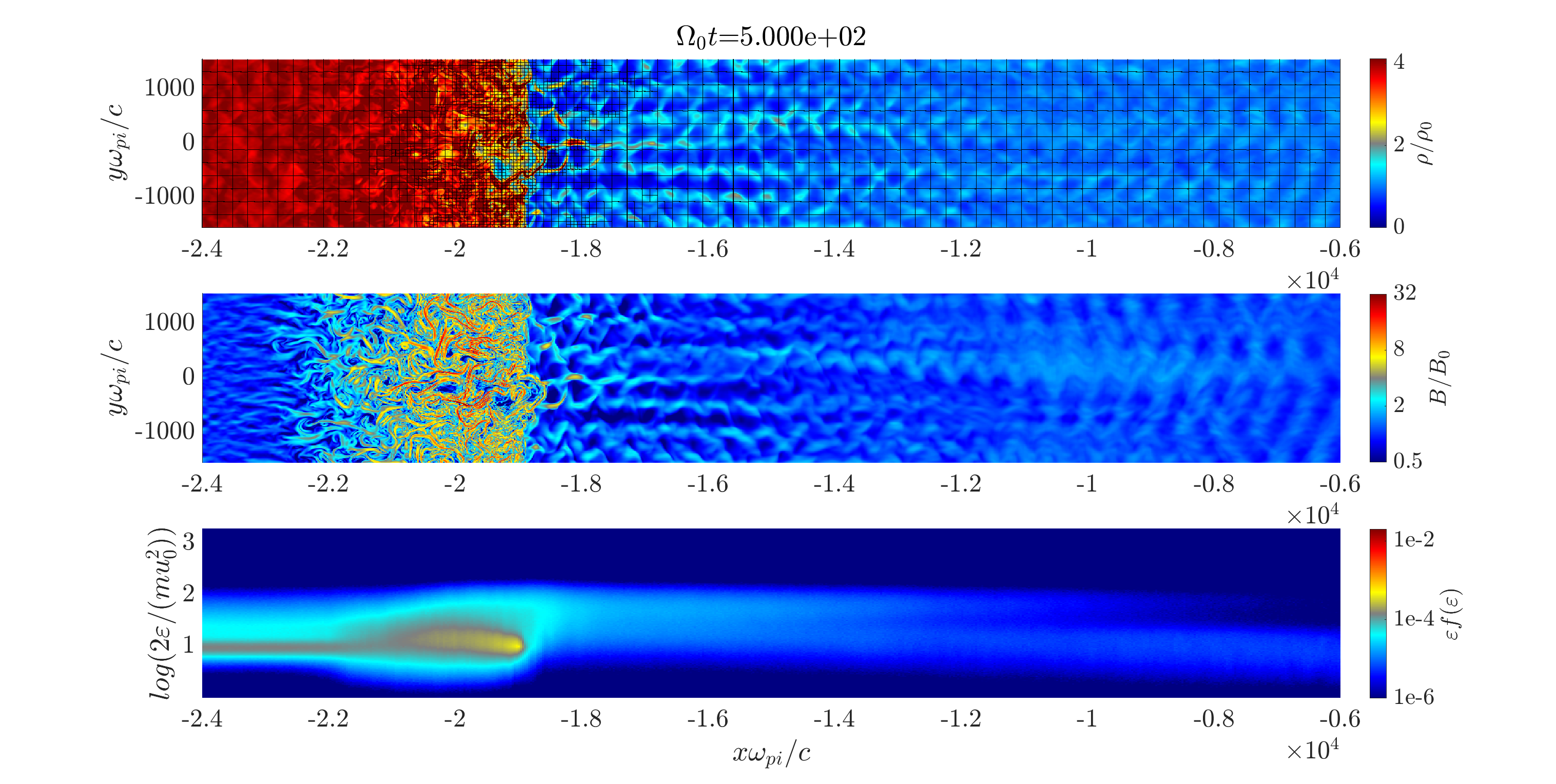}
    \caption{A snapshot of gas density (top), magnetic field strength (middle), and the CR distribution (bottom), at $t=500 \Omega_0^{-1}$ in the fiducial shock simulation with AMR. Note that only the part of the simulation box near the shock is shown. The grid in the top panel marks the boundary between MeshBlocks, each containing $20\times 20$ cells. The color bar of the CR distribution function has been normalized by CR particle energy $p_{CR}^2/(2m)$.
        The animation from $t=10^2$ to $t=1.2\times 10^3\Omega_0^{-1}$ is available in the online version of the journal.}
    \label{fig::shock_amr}
\end{figure*}

\begin{figure}
	\centering
    \begin{subfigure}[]{\columnwidth}
    	\includegraphics[width=\columnwidth]{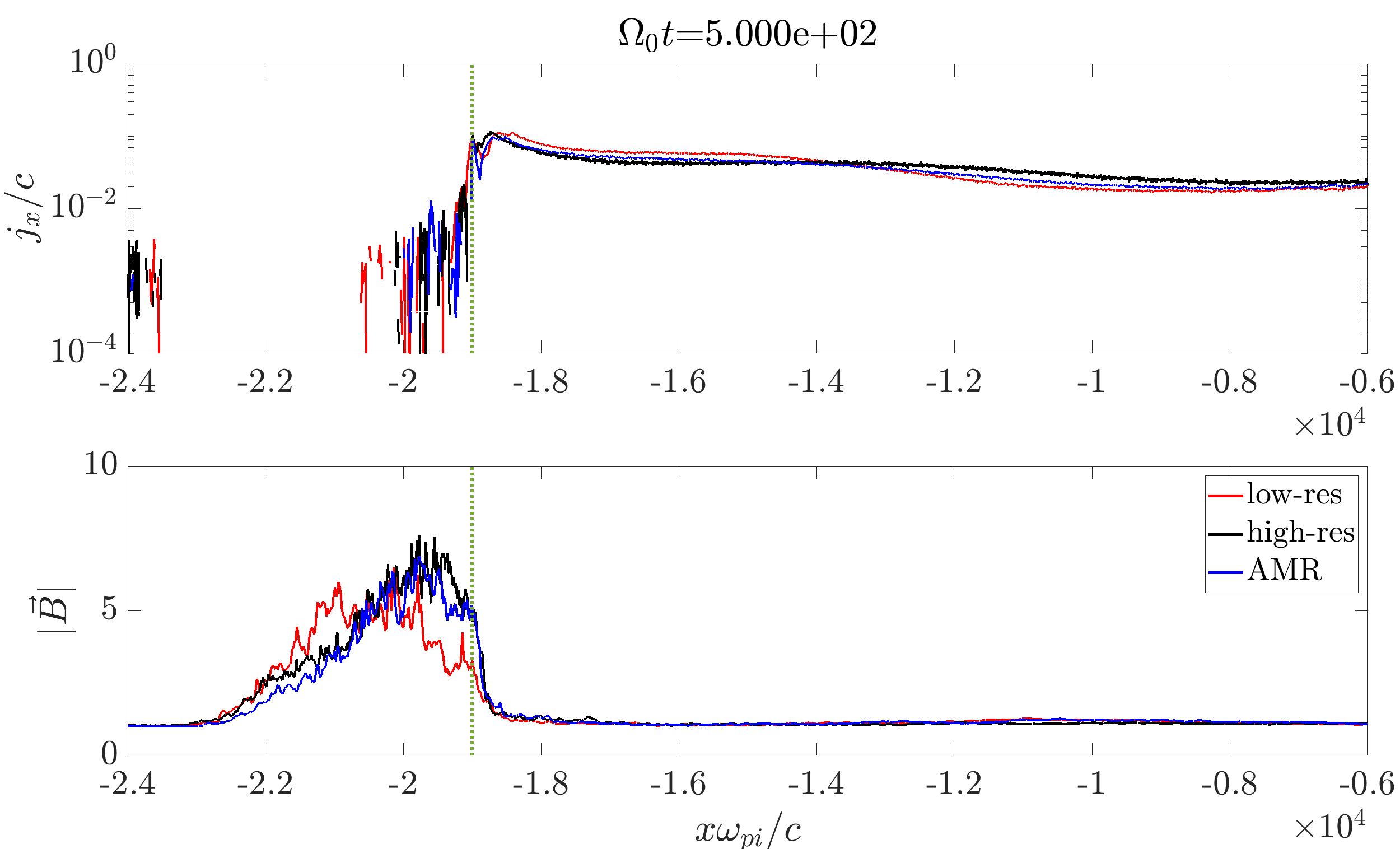}
        \caption{}
        \label{fig::shock_profile}
    \end{subfigure}
	\begin{subfigure}[]{\columnwidth}
    	\includegraphics[width=\columnwidth]{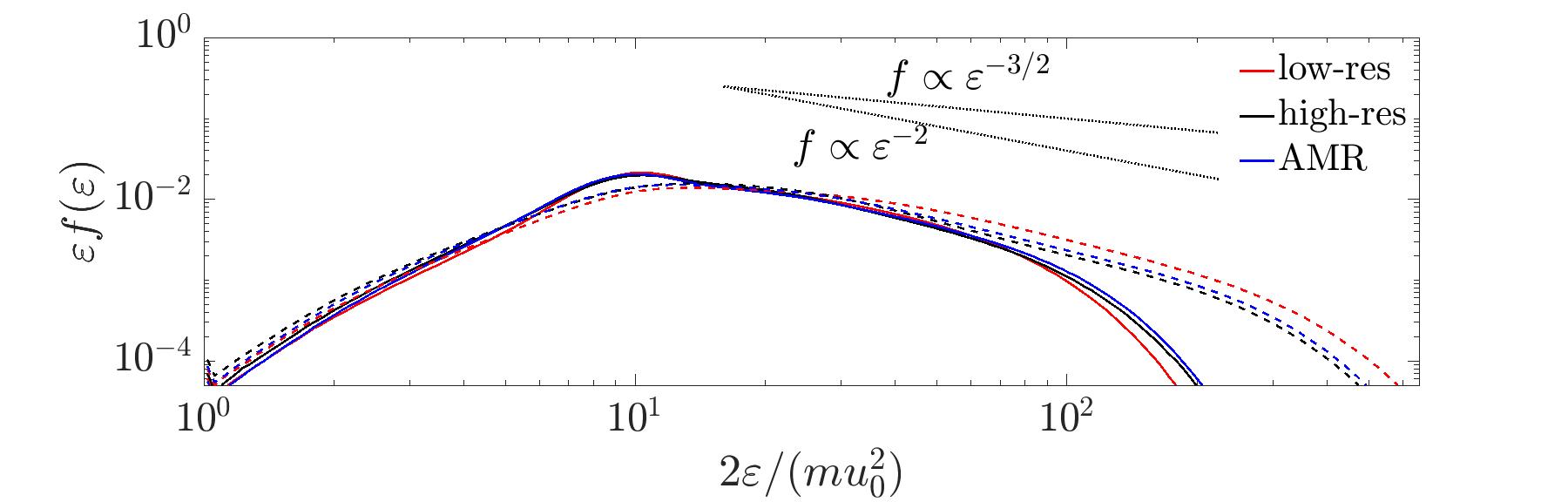}
        \caption{}
        \label{fig::shock_spec}
    \end{subfigure}
    \caption{Panel~\ref{fig::shock_profile}: The profiles of current density (top) and magnetic field strength (bottom) along the shock, averaged over the $y$-direction. 
    The green dotted lines mark the ideal shock location, also as the CR injection surface.
    Panel~\ref{fig::shock_spec}: the CR downstream momentum spectra (in dimensionless $\varepsilon f(\varepsilon)$). Colors indicate the runs with different resolutions. The solid lines are from time $t=500\Omega_0^{-1}$ and the dashed lines are from time $t=1200\Omega_0^{-1}$. Note that the CR spectrum in the AMR run largely overlaps the high resolution result.  The animation from $t=100$ to $1200\Omega_0^{-1}$ is available in the online version of the journal.}
\end{figure}

Next, we investigate the non-thermal proton acceleration in a non-relativistic shock using the MHD-PIC approach, which reproduce previous results while demonstrating the AMR capability of the code.

Protons can undergo Fermi acceleration \citep[e.g. ][]{1949PhRv...75.1169F, 1978ApJ...221L..29B, 1978MNRAS.182..147B} in magnetized collisionless shocks, where a small fraction of background thermal particles are accelerated to produce a non-thermal particle population, namely the CRs. PIC simulations (including hybrid PIC, e.g. \citealp{2011ApJ...726...75S, 2014ApJ...783...91C, 2015PhRvL.114h5003P, 2019MNRAS.485.5105C}) have simulated the proton acceleration process but 
with substantial
computational costs to resolve the micro-scales. By bypassing the micro-scales, MHD-PIC simulations \citep[e.g. ][]{2015ApJ...809...55B, 2018MNRAS.473.3394V, 2018ApJ...859...13M}
can dramatically reduce the numerical cost, and obtain the CR energy spectra similar to those in conventional PIC simulations, 
provided a proper injection prescription of supra-thermal particles at the shock front.

We employ a similar parallel shock setup as in \citet{2015ApJ...809...55B}. A uniform super-Alfv\'enic MHD flow along the $\hat{x}$-direction collides into a reflecting wall at the left $x$-boundary, with velocity $-u_0 \hat{x}$ and the background magnetic field $B_0 \hat{x}$. It forms a high-Mach number shock that propagates with velocity $\boldsymbol{u}_\text{sh}' = \left(\Gamma - 1\right) u_0 \hat{x}/2$ 
towards the right.
As CR injection is governed by microphysics processes not captured in MHD-PIC simulations, we artificially inject supra-thermal protons at the ideal shock surface $x=u_\text{sh}' t$, evenly and continuously \footnote{This treatment is simpler than previous works \citep[e.g. ][]{2015ApJ...809...55B, 2018MNRAS.473.3394V, 2018ApJ...859...13M} which dynamically track shock motion. It suffices here for test purposes where we only follow the shock for relative short time.}. The amount of injected protons is equal to the mass that the shock sweeps multiplied by the injection efficiency $\eta$. The injected protons are mono-energetic and isotropic relative to the ideal shock surface,
with $p/m = \sqrt{10}u_0$, and we subtract mass and momentum (energy) of the injected particles from local thermal gas.

The simulation parameters are similar to those adopted in \citet{2015ApJ...809...55B}. We use a 2D simulation box with  $L_x\times L_y = \left(48 \times 3.12\right)\times 10^3 c/\omega_\text{pi}$, where $c/\omega_\text{pi} \equiv \left(q/mc\right)^{-1}/ \sqrt{\rho_0}=U_{A0}/\Omega_0$ is the upstream ion inertial length, $\Omega_0\equiv B_0 q /mc$ is the cyclotron frequency in background field, and $U_\text{A0} \equiv B_0 /\sqrt{\rho_0}$ is the background Alfv\'en speed. The $y$-direction is periodic and is large enough to accommodate several gyro radii of high energy CRs. We set a high initial Alfv\'en Mach number $M_\text{A} \equiv u_0 / U_\text{A0}= 30$, the ideal gas equation of state with $\Gamma=5/3$ and a constant injection efficiency $\eta = 10^{-3}$. The numerical speed of light is set to $\mathbb{C} = 10^4 U_\text{A0}$, much greater than the shock velocity, so that particles will stay in the non-relativistic limit. Additionally, to minimize the influence of the initial state, we remove the particles injected before $45 \Omega_0^{-1}$ which largely travel freely into the unperturbed upstream. The computational units here are the upstream ion inertial length $c/\omega_\text{pi}$, the initial upstream gas density $\rho_0$, the parallel unperturbed magnetic field $B_0$, the the cyclotron frequency $\Omega_0$. 

We enable the AMR in the fiducial run which has two levels of refinement (3 levels in total), and compare the
results with two other uniform-grid runs with the same physical parameters but different resolutions. The cell size in the fiducial run ranges from $12 c/\omega_\text{pi}$ at the root level, 
and decreases by factors of 2 down to $3c/\omega_\text{pi}$ at the finest level. We adopt the following refinement criteria which depends on the density curvature $g_\rho$ and the pressure curvature $g_P$ \citep{2020ApJS..249....4S},
\begin{align*}
	g_{\rho, i, j} = |\frac{\rho_{i+1,j}}{\rho_{i,j}} + \frac{\rho_{i-1,j}}{\rho_{i,j}} - 2| + |\frac{\rho_{i,j+1}}{\rho_{i,j}} + \frac{\rho_{i,j-1}}{\rho_{i,j}} - 2|, \\
	g_{P, i, j} = |\frac{P_{i+1,j}}{P_{i,j}} + \frac{P_{i-1,j}}{P_{i,j}} - 2| + |\frac{P_{i,j+1}}{P_{i,j}} + \frac{P_{i,j-1}}{P_{i,j}} - 2|,
\end{align*}
where $i$ and $j$ are cell indices in the $x$ and $y$ directions. The program will refine the cells once the maximum $g_\rho$ or $g_P$ in one MeshBlock exceeds 1.0.
Conversely, if all active cells satisfy $g_\rho < 0.1$ and $g_P < 0.1$, the program will derefine this MeshBlock. The other two runs have uniform grids, whose resolutions correspond to
the coarsest and finest resolutions of the fiducial run. Regarding the number of simulation particles, we set the individual particle mass such that the downstream should hold $\sim 40$ppc in the finest cells and $\sim 640$ppc in the coarsest ones to achieve the desired injection efficiency $\eta$.

We show a snapshot of the early shock evolution and magnetic field amplification of the fiducial run in Figures~\ref{fig::shock_amr}~\&~\ref{fig::shock_profile}. The corrugated shock surface, as well as filaments and cavities in the upstream gas density (top panel in Figure~\ref{fig::shock_amr}) are characteristic of
the development of the Bell instability \citep{2005MNRAS.358..181B,2012MNRAS.419.2433R}.
The Bell instability is triggered by the upstream CR current from the injected CRs, which can be seen in Figure \ref{fig::shock_profile}.
Compared to the initial background field, the upstream field strength is amplified by a factor of up to $\sim$2-4 in certain regions (middle panel in Figure~\ref{fig::shock_amr}). Similar results in the shock structure and the magnetic field amplification are present in the other two uniform grid runs, where the fiducial run behaves much more close to the run in high resolution while slightly deviates from the low resolution run. The results are also broadly consistent with the previous simulation \citep[e.g. ][]{2015ApJ...809...55B, 2018ApJ...859...13M, 2018MNRAS.473.3394V}.

The injected protons are scattered by turbulent magnetic field and are accelerated to higher energies via the diffusion shock acceleration (DSA) process. Those high energy protons that exit the DSA process are then left downstream.
We sum up the simulation particles in the downstream to obtain the CR energy distribution $f(\varepsilon)$ in Figure~\ref{fig::shock_spec}. The maximum energy of non-thermal particles increases with time, and meanwhile the non-thermal particles form a power-law like tail in the CR spectrum. 
The power law is initially steep (e.g., at time $t=500\Omega_0^{-1}$), but later approaches the desired slope of $f(\varepsilon)\propto \varepsilon^{-3/2}$ (e.g., at time $t=1200\Omega_0^{-1}$).

\begin{figure*}
	\includegraphics[width=\textwidth]{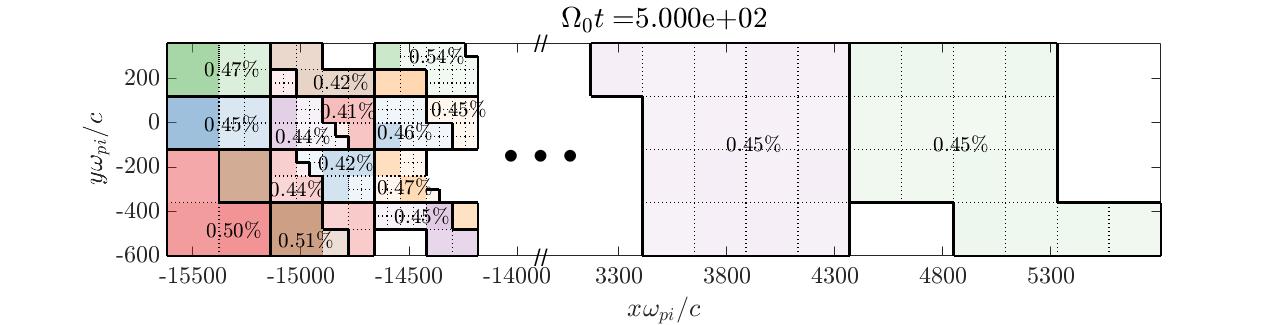}
    \caption{A portion of the load balancing workload in the AMR run at $t=500\Omega_0^{-1}$. The color box outlined by solid lines refers to the region that one process works on. Only 16 processes, out of the total 224 processes, are presented in the panel. The workload percentage for each processes is given in the number, and that for each MeshBlock, whose boundary is dashed lines, is given in the opacity.}
    \label{fig::load_balancing_snapshot}
\end{figure*}

In this shock problem, the spatial distribution of the CRs is highly inhomogeneous, with most CRs staying near the shock surface and downstream. As a result, the workload of individual MeshBlock is uneven.
We apply the load balancing to our fiducial run and plot the assignments of 16 processes (out of 224 in total) at time $t=500\Omega_0^{-1}$ in Figure~\ref{fig::load_balancing_snapshot}. Among them, 14 processes handle the MeshBlocks in the downstream, where the MeshBlocks are typically in the finer level with more particles per MeshBlock.
Two process owns a large number of 33 MeshBlocks in the upstream, as CR particles in the upstream are sparse and more workload there spends on the MHD side. We see that although individual processes own different number of MeshBlocks, the overall assignment of workload is nearly even, which saves the overall runtime.

\subsection{The CR gyro-resonant instability}
\label{sec::test_gyro-resonance}

We conduct the CR gyro-resonant instabilities to test our implementation of the $\delta f$ method and the expanding box module.
The CR gyro-resonant instability arises when the CR distribution deviates from isotropic by more than $U_A/c$ \citep{2005ppa..book.....K}, where $U_A$ is the local Alfv\'en velocity, which excites certain branches of the Alfv\'en modes by the resonant interactions the CRs. Given that typically $U_A\ll c$, capturing the CR gyro-resonant instabilities requires delicate representation of the angular distribution of the CRs that can accurately capture small deviations from isotropy, demanding the use of the $\delta f$ method \citep{2019ApJ...876...60B} to suppress Poisson noise.

The CR gyro-resonant instabilities come with two main flavors,
the CR streaming instability (CRSI, \citealp{1969ApJ...156..445K}) and the CR pressure anisotropy instability \citep[CRPAI ][]{2006MNRAS.373.1195L, 2020ApJ...890...67Z, 2018MNRAS.476.2779L}. The CRSI occurs when CRs' bulk drift (streaming) speed along the magnetic field exceeds $U_A$ in the gas co-moving frame, and it excites forward propagating, both left and right polarized Alfv\'en waves. The Bell instability (see Section~\ref{sec::test_bell}) can be considered as an extreme version of the CRSI with highly super-Alfv\'enic streaming, where the CR current dominates over gyro-resonance to primarily trigger the growth of the right-polarized Alfv\'en mode.
The CRPAI arises from the CR anisotropy in the momentum space whose the anisotropy level is greater than $\sim U_A/c$. The prolate pitch angle distribution and the oblate distribution excite both forward and backward propagating, but opposite circularly-polarized Alfv\'en modes. Since the CR anisotropy can be triggered by expansion/compression transverse to magnetic field, we can also test our expanding box implementation with the CRPAI.

\subsubsection{The CR streaming instability}
\label{sec::test_streaming}

We set up an isotropic homogeneous  $\kappa$ distribution for CRs where background gas drifts at speed $-v_d \hat{x}$ in a 1D periodic simulation box, same as \citet{2019ApJ...876...60B}. The CR distribution is characterized by the power index $\kappa$, the CR spatial number density $n_\text{CR}$ and the peak momentum $p_0$,
\begin{equation}
    f_{0,\text{iso}}\left(\boldsymbol{p}\right) = \frac{n_\text{CR}}{\left(\pi \kappa p_0^2\right)^{1.5}} \frac{\mathcal{G}\left(\kappa + 1\right)}{\mathcal{G}\left(\kappa - 0.5 \right)}\left(1 + \frac{p^2}{\kappa p_0^2}\right)^{-\kappa - 1}, \label{equ::kappa_dist_iso}
\end{equation}
where  $\mathcal{G}()$ refers to the Gamma function.
The gas
is initialized with a uniform density $\rho_0$ and 
background field $B_0 \hat{x}$. 
The dispersion relation
for left-polarized ($+$) and right-polarized ($-$) Alfv\'en waves is given by
\begin{align}
	\omega^2 = k^2 U_A^2 & \mp \frac{m n_\text{CR}}{\rho_0} \Omega_0 \omega \left(1 - \frac{k v_d}{\omega} \right) \left(1 - Q_1 \pm i Q_2 \right), \label{equ::crsi_dispersion} \\
	Q_1\left(k\right) \equiv& \frac{2}{\sqrt{\pi}\kappa^{1.5}}\frac{\mathcal{G}\left(\kappa + 1\right)}{\mathcal{G}\left(\kappa - 0.5 \right)} \left(\frac{m\Omega_0}{k p_0}\right)^3 \notag \\ &\times  \int_0^\infty \ln\left| \frac{1+s}{1-s}\right| \left(1 + \frac{s^2 m^2\Omega_0^2}{\kappa k^2 p_0^2}\right)^{-\kappa - 1} s \dd s, \\
	Q_2\left(k\right) \equiv& \frac{\sqrt{\pi}}{\kappa^{1.5}}\frac{\mathcal{G}\left(\kappa + 1\right)}{\mathcal{G}\left(\kappa - 0.5 \right)} \frac{m\Omega_0}{k p_0} \left(1 + \frac{m^2\Omega_0^2}{\kappa k^2 p_0^2}\right)^{-\kappa}, \label{equ::gyro_resonance_q}
\end{align}
and $\Omega_0$ is the cyclotron frequency.
In the limit of $mn_\text{CR}/\rho_0\ll m U_A^2/(v_dp_0)$, which is usually applicable in the ISM, the
instability growth rate (the imaginary part of $\omega$) is simplified as
\begin{equation*}
	\text{Im}\left(\omega\right) \approx - \frac{m n_\text{CR}}{2 \rho_0} \Omega_0 \left(1 - \frac{v_d}{U_A} \text{sgn}\left(k\right) \right) Q_2 \left(k\right).
\end{equation*}
The wave with wavenumber $k$ resonates with the CRs whose momenta $p_x/m \approx k/\Omega_0$. The instability occurs when $v_d>U_A$, and the fastest growing wavenumber $k_0$ corresponds to the waves that resonate with $p_x=p_0$.

The simulation parameters closely follow those in the fiducial run of \cite{2019ApJ...876...60B}. We set $v_d = 2 U_A$, $m n_\text{CR}/\rho_0 = 10^{-4}$,
$\mathbb{C}=300 U_A$,
$p_0 = m \mathbb{C}$, and 
$\kappa=1.25$.
We use an isothermal equation of state with a uniform sound speed $c_s = U_A$.
The most unstable wavelength is thus $2\pi p_0/\Omega_0m\approx1885U_A/\Omega_0$.
The simulation box is $L_x=96000 U_A/\Omega_0$, long enough to accommodate multiple most unstable wavelengths,
with a resolution of $\Delta x = 10 U_A/\Omega_0$.
We additionally impose a spectrum of forward and background traveling, left and right polarized Alfv\'en waves in random phase, whose initial wave intensity is given by $I(k) = A^2 / |k|$.
The numerical units are $\rho_0=B_0=q/\left(mc\right)=\Omega_0=U_A=1$,
and we choose $A=10^{-3}$.
Also following \cite{2019ApJ...876...60B}, we initialize the particles whose momenta range from $p_0/500$ to $500 p_0$,
represented by multiple particle species divided by 8 logarithmic bins to further reduce the numerical noise,
with each bin (or particle species) having 256 particle per cell.
Particles from different bins share the same $q/\left(mc\right)$ but are weighted in different particle mass $m^\text{sim}$ 
according to the distribution function.

\begin{figure}
	\includegraphics[width=\columnwidth]{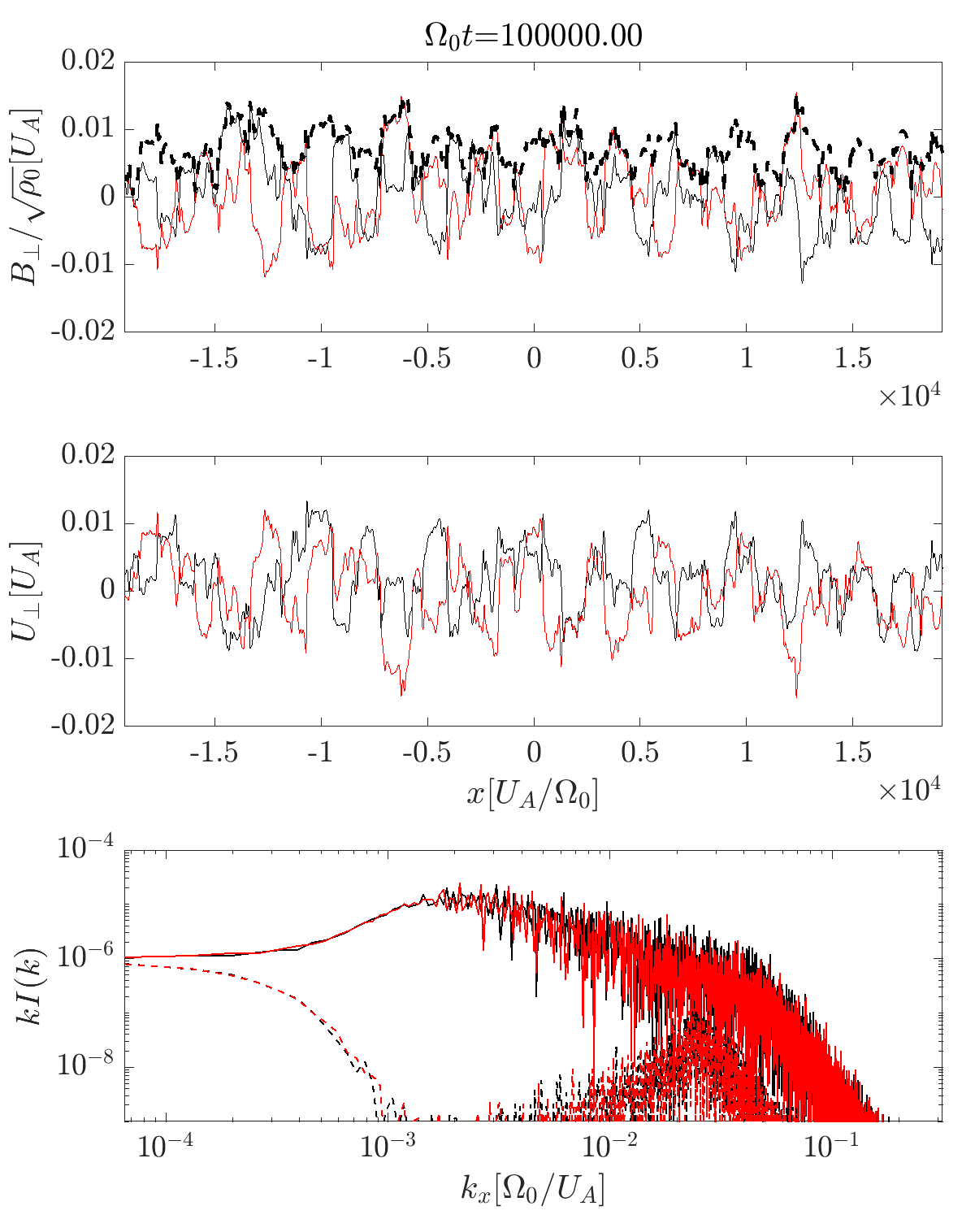}
    \caption{Snapshot at the relatively early evolutionary stage for the CRSI with the drift velocity $v_d = 2 U_\text{A}$, and $mn_{\rm CR}/\rho_0=10^{-4}$. Top and middle panels show the local profile of the perpendicular components of magnetic field and velocity (black and red solid lines for the y and z components). The black dashed line gives the total strength of the fluctuating field $|\boldsymbol{B}_\perp|$. The bottom panel shows the wave intensity spectra for forward (solid) and backward (dashed) propagating modes. Left/right handed branches are marked in black/red. The animation from $t=0$ to $1.4\times 10^5\Omega_0^{-1}$ is available in the online version of the journal.}
    \label{fig::crsi_profile}
\end{figure}

\begin{figure}
	\includegraphics[width=\columnwidth]{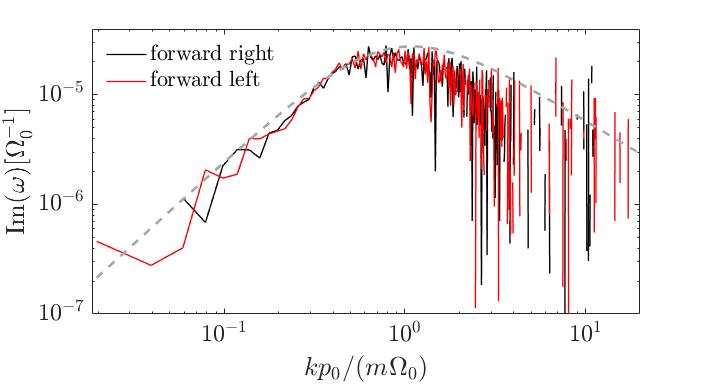}
    \caption{The measured linear growth rate of the CRSI with the drift velocity $v_d = 2 U_\text{A}$, and $mn_{\rm CR}/\rho_0=10^{-4}$, where black/red curves mark the right/left polarized forward waves. The gray dashed line mark the analytical growth rate expected from Equation~\ref{equ::crsi_dispersion}.}
    \label{fig::crsi_growth_rate}
\end{figure}

Figure~\ref{fig::crsi_profile} illustrates the relatively early evolutionary stage of the CRSI at $10^5\Omega_0^{-1}$. 
We follow the procedures outlined in \cite{2019ApJ...876...60B} to obtain the wave intensity spectra of each wave modes.
Forward-propagating waves grow and while backward-propagating waves are damped, and the wave spectrum peaks around $k\sim k_0$ (being the fastest growing mode), as expected.
The wave intensity spectrum extends to high-$k$, akin to the ubiquitous presence of rotational discontinuities as seen in the upper panels \citep{1974PhFl...17.2215C, 2021ApJ...914....3P}.
We further fit the wave amplitudes over the linear growth phase, from $t=0$ to $2\times 10^4\Omega_0^{-1}$, and the result is shown in
Figure~\ref{fig::crsi_growth_rate} on the $k$-by-$k$ bases. Overall, the results well agree with analytical dispersion relation, and are consistent with results obtained using the \texttt{Athena} code in \cite{2019ApJ...876...60B}.

\subsubsection{The CR pressure anisotropy instability}
\label{sec::test_anisotropy}

The setup for CRPAI is mostly identical to that for the CRSI, except that we initialize an anisotropic $\kappa$ distribution for the CRs and a static MHD background,
\begin{equation}
	f_{0,\text{aniso}}\left(\boldsymbol{p}\right) =\frac{n_\text{CR} \xi^2}{\left(\pi \kappa p_0^2\right)^{1.5}} \frac{\mathcal{G}\left(\kappa + 1\right)}{\mathcal{G}\left(\kappa - 0.5 \right)}\left(1 + \frac{p^2}{\kappa p_0^2}\left(\mu^2  - \xi^2 \mu^2 + \xi^2\right)\right)^{-\kappa - 1}, \label{equ::kappa_dist_aniso}
\end{equation}
where the cosine of pitch angle and the anisotropy level are $\mu$ and $|\xi^2 -1|$. When $\xi > 1$, CRs concentrate towards $\mu = 0$ and form an oblate distribution, while a prolate  distribution ($\xi < 1$) has more CRs concentrate towards $\mu=\pm 1$. In the non-relativistic regime, the analytic dispersion relation of CRPAI is given by
\begin{equation}
	\omega^2 = k^2 U_A^2 \mp \frac{m n_\text{CR}}{\rho_0} \Omega_0 \omega \left(1 \pm \frac{1-\xi^2}{\xi^2} \frac{\Omega_0}{\omega} \right) \left(1 - \frac{Q_1}{\xi^2} \pm i \frac{Q_2}{\xi^2}\right), \label{equ::crai_dispersion}
\end{equation}
where $Q_1$ and $Q_2$ are identical to Equation~\ref{equ::gyro_resonance_q}. \cite{2018MNRAS.476.2779L} numerically studied the CRPAI with $\xi^2 \sim 0.8$ or $1.2$ without employing the $\delta f$ method, where the simulations involve substantial numerical noise. Here, we consider $|\xi^2 -1|\sim 0.02\ll1$, likely much closer to the true level of (very low) anisotropy in the ISM.
In this limit, the instability growth approximates to
\begin{equation*}
	\text{Im}\left(\omega\right) \approx - \frac{m n_\text{CR}}{\rho_0} \Omega_0 \left(1 \pm \frac{1-\xi^2}{\xi^2} \frac{\Omega_0}{k U_A} \right) \frac{Q_2 \left(k\right)}{2\xi^2},
\end{equation*}
where $\pm$ is related to whether the mode is left or right polarized. The instability criteria is given by $\xi < \sqrt{\Omega_0 / \left(\Omega_0 + U_A k\right)}$ or $\xi> \sqrt{\Omega_0 / \left(\Omega_0 - U_A k\right)}$.
The anisotropic CR distribution generate left or right polarized waves that propagate in both directions, while the sense of polarization is the opposite between waves excited by a prolate or oblate distribution. Waves of the opposite polarization are damped.

To simulate the CRPAI in the linear phase and reproduce the analytic growth rate (Equation~\ref{equ::crai_dispersion}), we first increase the numerical speed of light to $\mathbb{C}= 3\times 10^4 U_A = 10^2 p_0/m$ to ensure that most particles are in the non-relativistic regime, while other parameters are mostly identical to those in Section~\ref{sec::test_streaming}. Note that, in the long wavelength limit, the instability growth rate is proportional to $k^{2\kappa - 2}$. We shift $\kappa$ to 1.75 to ensure $\text{Im}\left(\omega\right) \ll k U_A$ for all wavelength. Also,
the CRPAI excites both forward and backward waves,
and by symmetry, the CR distribution is expected to be always symmetric with respect to $90^\circ$ pitch angle.
Therefore, a relatively lower resolution is feasible without needing to care for the $90^\circ$ barrier, and we choose $\Delta x = 20 U_A/\Omega_0$. The initial anisotropy level is set as $\xi=1.01$ for an oblate CR distribution and $\xi=0.99$ for a prolate one, respectively.

\begin{figure}
	\includegraphics[width=\columnwidth]{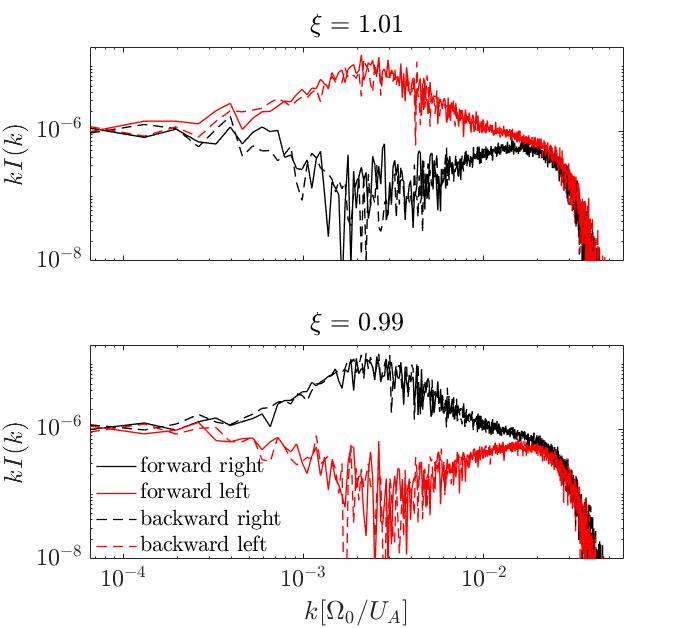}
    \caption{Wave intensity spectra, $kI(k)$, for CRPAI with the anisotropy parameter $\xi=1.01$ (top) and $0.99$ (bottom), at $t=6 \times 10^3\Omega_0^{-1}$. Forward and backward propagating waves are in solid and dashed lines. Left and right handed branches are marked in black and red respectively.}
    \label{fig::crai_once_spec}
\end{figure}
\begin{figure}
	\includegraphics[width=\columnwidth]{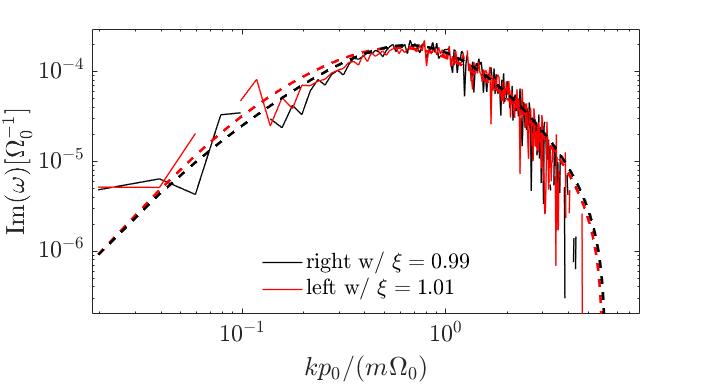}
    \caption{The measured linear growth rate of the CRPAI with $mn_{\rm CR}/\rho_0=10^{-4}$. The black curve refers to right polarized forward waves with the CR anisotropy parameter $\xi=0.99$, and the red curves refers to the left polarized forward waves with $\xi=1.01$. The dashed lines mark the analytical growth rate expected from Equation~\ref{equ::crai_dispersion}.}
    \label{fig::crai_growth_rate}
\end{figure}

We show the wave intensity spectra for polarized Alfv\'en waves in Figure~\ref{fig::crai_once_spec}, for the $\xi=1.01$ and $0.99$ cases. As expected, the oblate CR distribution ($\xi>1$) only amplifies right-polarized waves from the initial intensity spectrum $k I(k) = 10^{-6}$, while the prolate CR distribution ($\xi<1$) only amplifies left-polarized waves. 
We quantitatively measure the wave growth rates, shown
in Figure \ref{fig::crai_growth_rate}, and the results well coincide with the analytical solution from Equation~\ref{equ::crai_dispersion}.

\subsubsection{The CRPAI in an expanding/compressing box}

\begin{figure}
	\includegraphics[width=\columnwidth]{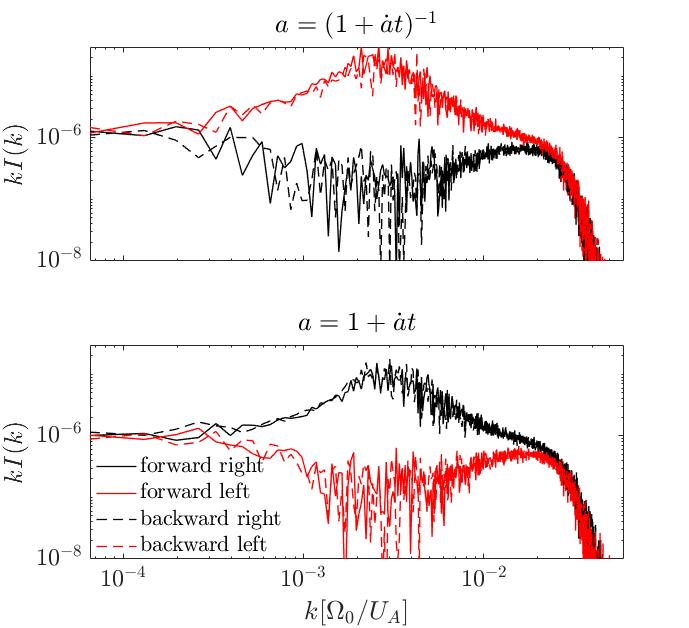}
    \caption{Same as Figure~\ref{fig::crai_once_spec}, but for compressing (top) and expanding (bottom) boxes, with the compressing/expanding rate $\dot{a}=1 \times 10^{-5} \Omega_0$ at $t=6\times 10^{3}\Omega_0^{-1}$.}
    \label{fig::crai_drive_spec}
\end{figure}

The previous simulations employ an initially anisotropic CR distribution in the non-relativistic limit. We next simulate CRPAI in a more realistic setting by continuously driving the CRPAI in the expanding box framework and covering both non-relativistic and relativistic particles.
The CR anisotropy arises due to the conservation of magnetic moment $p^2(1-\mu^2)/B$, where adiabatic compression or expansion of the background gas (e.g., in the ISM) generally leads to increase or decrease of background magnetic field strength \citep{2006MNRAS.373.1195L, 2020ApJ...890...67Z}. This simulation setup more naturally captures the fact that the gas in the ISM is always being compressed and/or rarefied by compressible turbulence at large scales, thus constantly driving CR anisotropy and hence trigger the CRPAI. We anticipate that the waves generated by the CRPAI represent a major complement to those by the CRSI, and play a significant role as the source of CR scattering that couples the CRs with background gas.

We initialize the simulation with an isotropic CR distribution (Equation~\ref{equ::kappa_dist_iso}),
set up an anisotropic expansion given by
$a_1 = a^2\left(t\right), a_2=a_3=a\left(t\right)$, 
so that an MHD wave can freely propagate along the background field in $x$-direction (Section~\ref{app::cpaw}). As $a_i p^i$ is conserved along background field, the resulting anisotropy parameter $\xi$ 
is expected to follow $a^{-1}(t)$ and increases/decreases with time in the compressing/expanding box. We carry out expanding/compressing box simulations with numerical parameters similar to those in Figure~\ref{fig::crai_once_spec}, using $\mathbb{C}=300 U_\text{A}$ and $\kappa=1.25$. The equilibrium CR distribution $f_0$ for the $\delta f$ method in the expanding/compressing boxes is equal to $f_{o, \text{iso}}\left(a_1 p_x, a_2 p_y, a_3 p_z\right)$ (See Section~\ref{app::gyration}). The simulation drives anisotropy slowly, where the expanding parameter $a(t)$ is set to $1+\dot{a}t$ or $(1+\dot{a}t)^{-1}$ with $\dot{a} = 10^{-5} \Omega_0$, so that driving would not distort CRs' gyro motion. 

We plot the simulation intensity spectra in Figure~\ref{fig::crai_drive_spec}. The wave spectra in the expanding/compressing boxes resemble those in Figure~\ref{fig::crai_once_spec}. 
With constant driving, the anisotropy parameter continuously increases/decreases so that the instability growth rate increases with time. The CRPAI then amplifies the same polarization branches as the case with an initially anisotropic CR distribution. The shape of the wave intensity spectrum is slightly modified, partly due to continuous driving, and partly due to our switch to a trans-relativistic particle distribution.
Note that without the CRs, the expanding/compressing can directly damp/amplify MHD waves (Equation~\ref{equ::app_solution_cpaw}) due to the adiabatic cooling/heating. As a result,
the expanding box gives weaker wave power in Figure~\ref{fig::crai_drive_spec}, compared with that in the compressing box. As the anisotropy level $|\xi^2 - 1| \approx 2 \dot{a}t$, the instability growth rate varies with time, and hence we do not pursue to fit for a growth rate.

\begin{figure*}
	\includegraphics[width=\textwidth]{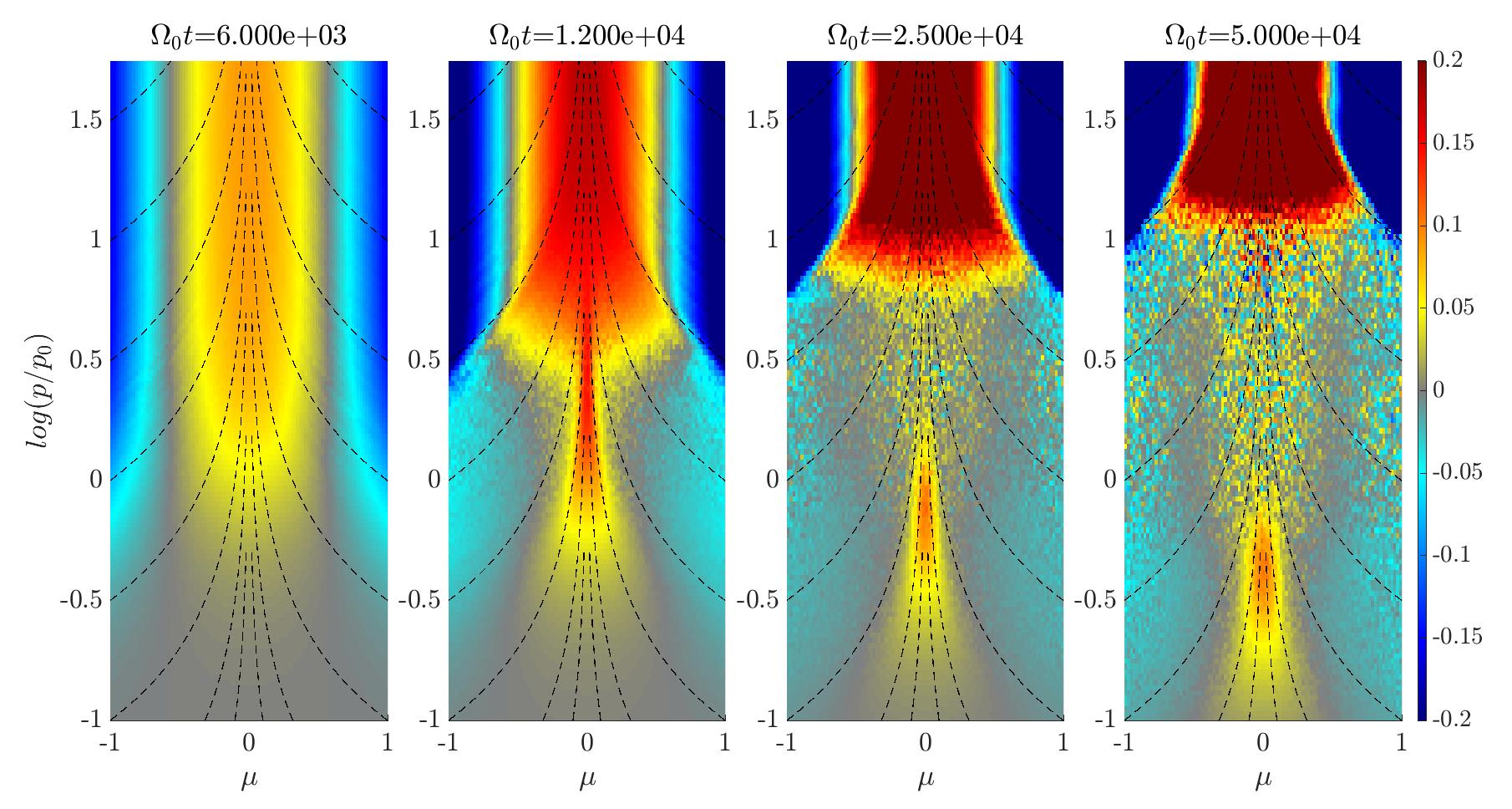}
    \caption{The CR distribution function $\left(f / \left\langle f\right\rangle_\mu - 1 \right)$ for the expanding box simulation ($a = 1 + \dot{a} t, \dot{a} = 1 \times 10^{-5} \Omega_0$) at a series of snapshots. The dashed lines refer to particle momenta that resonate with the same wave with constant $k=\Omega_0 m / \left(p \mu\right)$. The animation from $t=0$ to $5 \times 10^{4}\Omega_0^{-1}$ is available in the online version of the journal.}
    \label{fig::expand_cr_spec}
\end{figure*}
\begin{figure*}
	\includegraphics[width=\textwidth]{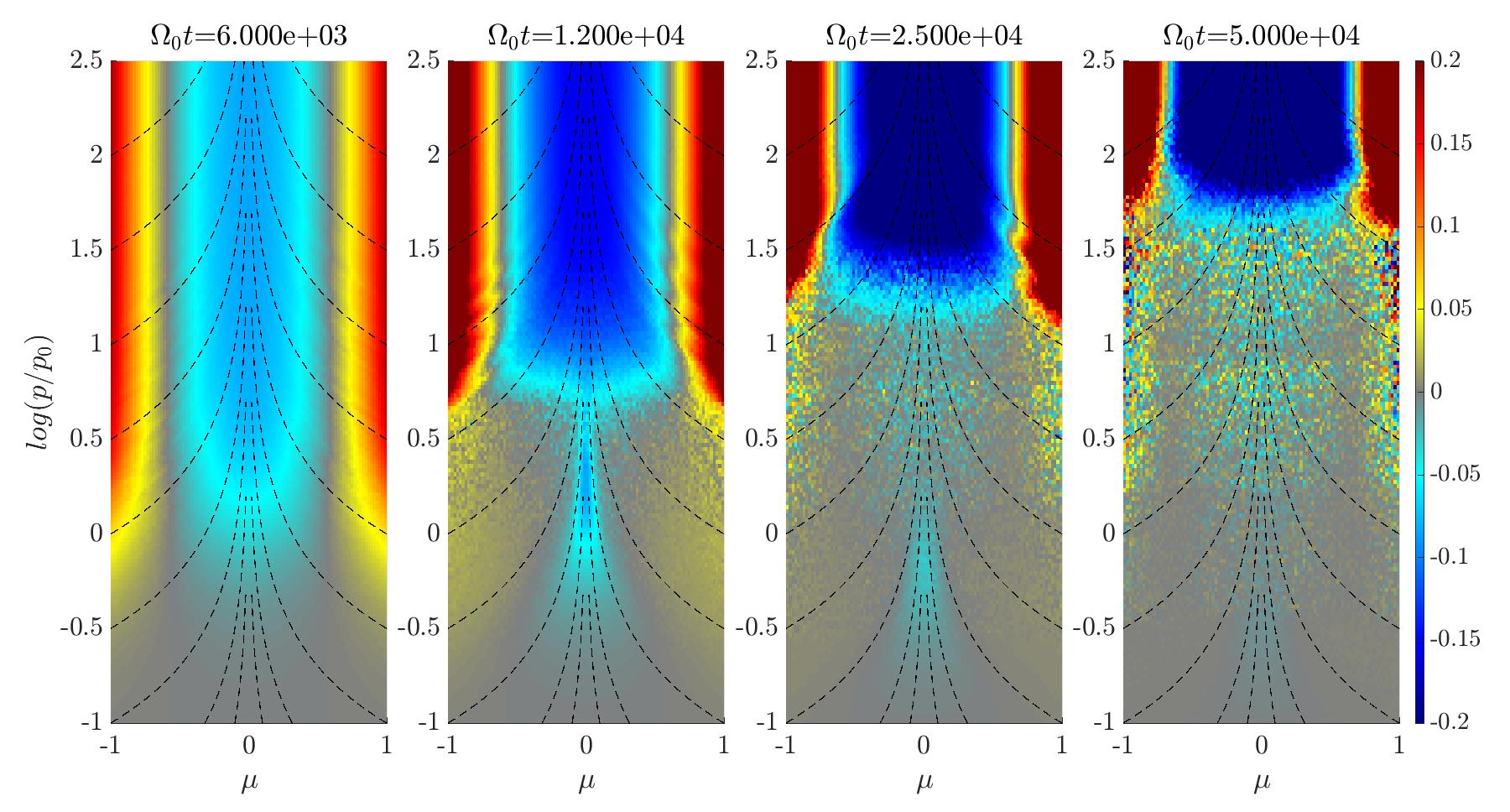}
    \caption{Same as Figure~\ref{fig::expand_cr_spec}, but for the compressing box simulation ($a = \left((1 + \dot{a} t\right)^{-1}, \dot{a} = 1 \times 10^{-5} \Omega_0$). The animation from $t=0$ to $5 \times 10^{4}\Omega_0^{-1}$ is available in the online version of the journal.}
    \label{fig::shrink_cr_spec}
\end{figure*}
In Figures~\ref{fig::expand_cr_spec}~\&~\ref{fig::shrink_cr_spec}, we compare the  CR distribution $f \left(p, \mu\right)$ and the isotropic distribution $\left\langle f\right\rangle_\mu\left(p\right)$ averaged over the pitch angle. Initially, before $\sim 1\times 10^4\Omega_0^{-1}$, the box's expansion/compression mainly drives CR distribution to build up anisotropy and the Alfv\'en waves are too weak to isotropize $f$. In an expanding box ($a=1+\dot{a}t$), more CRs concentrate around $\mu \sim 0$ and then trigger the CRPAI with an oblate $f$. With the right-polarized waves growing to sufficient amplitude, the CRs undergo the quasi-linear diffusion, resonating with the waves whose $k=\Omega_0 m / \left(p \mu\right)$ (black dashed lines in Figure~\ref{fig::expand_cr_spec}~\&~\ref{fig::shrink_cr_spec}). Since the CR scattering rate is proportional to the wave intensity $kI(k)$, isotropization is the most effective for particles with $p \sim p_0$, which nearly recovers isotropy. The isotropization for the particles with higher or lower momentum requires more time, after the intensity of the resonating waves increases sufficiently
to balance the driving of CR pressure anisotropy. 
For the same reason, the waves in the compressing box, with larger wave amplitudes, isotropizes $f$ more efficiently than that in the expanding box.
By the end of our simulations, the CRs around momentum $0.5-10p_0$ (expanding box) or $0.3-40p_0$ (compressing box) 
achieve saturated state, isotropization by wave scattering is balanced by the anisotropy driving by background expansion/compression. The CRs maintain certain weak anisotropy level, leading to another balance where wave growth by the CRPAI is balanced by wave damping (here mainly numerical). 

Following the success of this test that drives the CRPAI in a realistic manner, the natural next step is to implement certain wave damping mechanisms, so that wave growth and damping can be balanced in a controlled manner, which will allow us to measure the CR transport coefficient similar to that done in \citet{2022ApJ...928..112B} in the case of the CRSI. This will be explored in our future work.

\section{Summary and discussion}

This paper mainly describes the MHD-PIC module in \texttt{Athena++}. The  MHD-PIC method treats thermal ions and electrons as MHD gas, and CRs are represented by individual super particles. The MHD gas exerts the Lorentz force to CR particles, via the electromagnetic field, and the CR backreaction acts on the gas momentum (energy). The MHD-PIC method does not need to resolve the ion/electron skin depth and retains all CR kinetic physics, which substantially alleviates the issue of scale separation encountered in conventional PIC methods in CR-related kinetic problems. We implement the Boris pusher to numerically integrate CRs 
combined with the van Leer integrator for the MHD gas including the CR backreaction. The numerical scheme ensures the second-order accuracy and conserves the total momentum (energy).

We optimize the the MHD-PIC module for better vectorization efficiency and parallel performance. The vectorized interpolation and deposit requires efficient data access. We create the intermediate arrays so that each simulation particle can continuously access its neighboring multidimensional grid points (Figure~\ref{fig::interm}).  Dynamically sorting the simulation particle data also improves the efficiency of the interpolation and the deposit steps. Both the intermediate arrays and the dynamic sorting bring extra computational costs, but they enhance the overall performance. The cost to integrate one particle is equal to half of cost to update one MHD cell in 1D simulation, one forth of that in 2D and one sixth of that in 3D (Figure~\ref{fig::performance}). Meanwhile, the MHD-PIC module inherits the hybrid parallelization strategy and the dynamic scheduling from \texttt{Athena++}, and achieves an excellent parallel scaling (Figure~\ref{fig::weakscale}). The load balancing of the MHD-PIC module averages workload to all processors and makes the most of the given computational resource on distributed memory parallel systems.

We also implement additional techniques for specific applications. For the CRSI and CRPAI, the $\delta f$ method is essential to suppress the numerical noise when the parameters approach to the realistic regime. We implement the $\delta f$ method that is compatible with arbitrary background CR distributions. Meanwhile, we formulate and implement the "expanding box" framework to study the CRPAI as a (more realistic) driving problem.. The box expansion/compression continuously drives the CR pressure anisotropy which competes with the isotropization process arising from quasi-linear diffusion. To study the CR acceleration in shocks, the MHD-PIC module supports both AMR and SMR. The deposit scheme ensures smoothness across refinement boundaries (currently) in the Cartesian grid.

We examine the implementation with several benchmark tests, while each test corresponds to certain specific features in the code. The test particle gyration validates the accuracy of the Boris pusher (Section~\ref{sec::gyro}). After turning on the CR backreaction, we successively reproduce the linear growth rate of the Bell instability (Section~\ref{sec::test_bell}) and the gas oscillation with electrons’ and positrons’ gyration (Section~\ref{sec::rel_vel}) in the simulations. The SMR run and the AMR run for the electrons’ and positrons’ gyration test both converge to the same result in a uniform grid (Figure~\ref{fig::rel_vel}),
validating the interpolation/deposit scheme across refinement boundaries.
The result of CR acceleration in non-relativistic parallel shocks readily reproduces earlier works \citep[e.g. ][]{2015ApJ...809...55B, 2018MNRAS.473.3394V, 2018ApJ...859...13M}, where the upstream turbulence and magnetic field amplification is triggered by the Bell instability
and lead to the Fermi acceleration of the CRs that form a power law spectrum (Section~\ref{sec::test_shock}). Load balancing in \texttt{Athena++} assigns more processors to integrate the region around the shock front (Figure~\ref{fig::load_balancing_snapshot}).
We finally employ the $\delta f$ method to reproduce the linear growth rates of both the CRSI and the non-relativistic CRPAI.

Two astrophysical applications we mainly consider in this work are the particle acceleration in non-relativistic parallel shocks and the CR gyro-resonant instabilities. In the shock problem, as the MHD-PIC method cannot capture the initial process of particle energization from the thermal particle pool, the simulation needs a recipe for injecting supra-thermal particles. As a test problem, 
we neglect the corrugation of shock iso-surface and inject supra-thermal particles uniformly at the ideal shock front.
With more carefully designed injection recipes,
\citet{2015ApJ...809...55B, 2018ApJ...859...13M, 2018MNRAS.473.3394V}  have carried out longer-term simulations of non-relativistic parallel shocks and demonstrated efficient ion acceleration with the CR energy distribution matching theoretical expectations, and the results have been further extended to oblique shocks \citep{2018MNRAS.473.3394V, 2022ApJ...929....7V}.
The main next steps would be to further improve the injection recipes that better mimics the physics identified from kinetic simulations (e.g. \citealp{2013ApJ...773..158G, 2015ApJ...798L..28C}). This would also enable carrying out much longer term simulations to study how the shock properties is modified by CR acceleration \citep[][]{2015ApJ...809...55B, 2020ApJ...905....1H, 2020ApJ...905....2C}. Another direction would be to further extend the MHD-PIC method to the relativistic MHD regime, allowing us to investigate particle acceleration in relativistic shocks \citep[e.g. ][]{2011ApJ...726...75S}.

In Section~\ref{sec::test_anisotropy}, we studied two flavors of the CR gyro-resonant instabilities, the CRSI and CRPAI. The simulations of the CRSI follows the recent works conducted in the original \texttt{Athena} code \citep{2019ApJ...876...60B}, where the latter is not well studied in the literature \citep{2006MNRAS.373.1195L,2018MNRAS.476.2779L,2020ApJ...890...67Z}. These instabilities exploit the free energy of CR streaming and anisotropy that drive the growth of MHD waves (primarily Alfv\'en waves) that resonate with the CR gyro motion, and are most effective for low-energy CRs peaking around GeV that carry the bulk of the CR energy density. They play a key role in setting the momentum and energy coupling between the CRs and background gas described by the CR transport coefficients, thus determine the efficiency of CR feedback at macroscopic scales. 
In particular, we apply the expanding/compressing box framework to study the CRPAI as a driving problem, where a steady anisotropic CR distribution (Figure~\ref{fig::expand_cr_spec}~\&~\ref{fig::shrink_cr_spec}) is achieved by the competition between the expanding/compressing box driving anisotropy and the CR scattering by the Alfv\'en waves.
Currently, wave growth by the CRPAI is balanced by numerical dissipation. In the next step, we will further incorporate a number of physical wave damping mechanisms
(e.g. the ion-neutral damping, the non-linear Landau damping), and will aim to measure the CR
transport coefficients
from the first principles, analogous to the work of \citet{2022ApJ...928..112B} that applied the streaming box framework to measure the CR transport coefficients under the CRSI. With more efficient implementation, it will also allow us to extend the current 1D simulations to multi-dimensions,
where the fast mode waves are expected to simultaneously grow and to effectively isotropizes the CR distribution.
Eventually, efforts along this line will provide reliable microphysics-calibrated subgrid models of CR transport coefficients that will help eliminate major uncertainties in understanding CR transport and feedback.

The MHD-PIC method also finds applications in a variety of other astrophysical and plasma problems. For instance,
relativistic jets can trigger
MHD instabilities, which likely further leads to efficient particle acceleration towards ultra-high-energy CRs (UHECRs) \citep{2018PhRvL.121x5101A, 2020ApJ...896L..31D, 2022ApJ...931..137O, 2020arXiv200904158M, 2021ApJ...907L..44S, 2022ApJ...928...62K}. The jet may also re-accelerate existing CRs towards UHECRs under the ``espresso'' scenario \citep{2019ApJ...886....8M, 2021ApJ...921...85M}.
Extension of the method to integrate particles under the guiding-center approximation may further alleviate scale separation (when gyro-resonance is unimportant) \citep{2019PhPl...26j2903A, 2023CoPhC.28508625M}, 
potentially allowing the study of particle acceleration during magnetic reconnection.
In addition, there are also potential applications in fusion plasmas, where energetic fast particles can drive real-space instabilities (see \citealp{2016RvMP...88a5008C} and references therein).
We will make our MHD-PIC module in \texttt{Athena++} publicly available in the near future, and hope this module will benefit the broad plasma (astro)physics community on kinetic problems associated with energetic particles.

\section*{Acknowledgements}
We are indebted to Chao-Chin Yang and his team for sharing us an early version of the (dust-)particle module in \texttt{Athena++}. We thank Lev Arzamasskiy, Lin Gan, Qingfeng Li and Jianyuan Xiao for useful discussions. This research was supported by NSFC grant 11873033. Numerical simulations are conducted on TianHe-1 (A) at National Supercomputer Center in Tianjin, China, and on the Orion cluster at Department of Astronomy, Tsinghua University.

\section*{Data Availability}

The data underlying this article will be shared on reasonable request to the corresponding authors.



\bibliographystyle{mnras}
\bibliography{example} 




\appendix

\section{Benchmark tests for the expanding coordinates}
\label{app::test_expanding}

To verify our implementation of the expanding box source terms, we design two simple problems to separately verify the MHD solver and the CR particle integrator in the expanding/compressing boxes. In these tests, we have computed the analytical solutions to compare with simulation results.

\subsection{Circularly polarized Alfv\'en waves}
\label{app::cpaw}
We consider an Alfv\'en wave propagating in an expanding or compressing box. The background field $\boldsymbol{B}_g$, gas density $\rho$ and temperature profiles are all uniform. The gas velocity perturbation $\boldsymbol{u}_\perp$ and the magnetic field perturbation $\boldsymbol{B}_\perp$ follow the dynamical equations, simplified from Equation~\ref{equ::expand_mass_mhd}~to~\ref{equ::expand_field},
\begin{align}
    &\partial_t \left(l \rho \right) = 0, \\
	&\nabla' \cdot \boldsymbol{u}_\perp = 0, \\
    &\left(\partial_t + \boldsymbol{u}_\perp \cdot \nabla'\right) \boldsymbol{u}_\perp + \boldsymbol{u}_\perp \cdot \mathbb{D} + \nabla' \frac{b_\perp^2}{2} -\boldsymbol{b}_\perp \cdot \nabla' \boldsymbol{b}_\perp  = \boldsymbol{U}_\text{A} \cdot \nabla' \boldsymbol{b}_\perp, \\
    &\nabla' \cdot \boldsymbol{b}_\perp = 0, \\
    &\partial_t \boldsymbol{U}_\text{A} + \frac{\partial_t l}{2l} \boldsymbol{U}_\text{A} - \mathbb{D} \cdot \boldsymbol{U}_\text{A} = 0, \\
    &\partial_{t} \boldsymbol{b}_\perp + \frac{\partial_t l}{2l} \boldsymbol{b}_\perp - \mathbb{D} \cdot \boldsymbol{b}_\perp - \nabla' \cross \left(\boldsymbol{u}_\perp \times\boldsymbol{b}_\perp \right) = \left(\boldsymbol{U}_\text{A} \cdot \nabla'\right) \boldsymbol{u}_\perp, \\
	&\boldsymbol{b}_\perp \equiv \frac{\boldsymbol{B}_\perp}{\sqrt{\rho}}, \quad \boldsymbol{U}_\text{A} \equiv \frac{\boldsymbol{B}_\text{g}}{\sqrt{\rho}}. \notag
\end{align}
We omit the energy equation as it is irrelevant to the Alfv\'en wave propagation. It is convenient to rewrite the above equations in the Elsasser variables, $\boldsymbol{w}_\pm \equiv \boldsymbol{u}_\perp \mp \boldsymbol{b}_\perp$, which naturally express the evolution of the  Alfv\'en waves
\begin{align}
    \left(\partial_t  \pm \boldsymbol{U}_\text{A} \cdot \nabla'\right) \boldsymbol{w}_\pm =& -\boldsymbol{w}_\mp \cdot \nabla' \boldsymbol{w}_\pm - \frac{1}{8} \nabla' \left(\boldsymbol{w}_+ - \boldsymbol{w}_-\right)^2 \notag \\ &- \left(\mathbb{D} - \frac{\partial_t l}{4l}\right) \boldsymbol{w}_\mp - \frac{\partial_t l}{4l} \boldsymbol{w}_\pm.
\end{align}
The last two terms on the RHS are the source terms from the expanding box, also present in local models of the solar wind \citep{1980JGR....85.1311H, 2015ApJ...811...50C}.
The third term represents the conversion between forward and backward propagating modes. The last term indicates the waves' damping/growth due to the box's expansion/compression. For test purposes, we adopt a special case here, 
where the background field is along the $x$-direction and $a_1 = a^2(t)$, $a_2 = a_3 = a(t)$. In this way, the Alfv\'en speed remains a constant and there is no conversion between forward and backward waves
\begin{equation*}
	\left(\mathbb{D} - \frac{\partial_t l}{4l}\right) \boldsymbol{w}_\pm = 0.
\end{equation*}
Moreover, when only one circularly-polarized wave propagates, $\nabla' \boldsymbol{w}_\pm^2 =0$, the dynamical equation becomes,
\begin{equation}
	\left(\partial_t  \pm \boldsymbol{U}_\text{A} \cdot \nabla'\right) \boldsymbol{w}_\pm + \frac{\partial_t a}{a} \boldsymbol{w}_\pm = 0,
\end{equation}
where the analytic solution is
\begin{align}
	\boldsymbol{w}_\pm\left(t,x\right) & \propto  a^{-1}\left(t\right) \exp\left(i \phi_\text{wave} \left(t\right) \mp i k x\right) \notag \\ & \phi_\text{wave} \left(t\right) = \left\{
\begin{aligned}
& - k U_\text{A} \dot{a}^{-1} \left(a^{-1}- 1\right), \text{if } a=1+\dot{a}t,\\
&  k U_\text{A} \dot{a}^{-1} \left(a^{-3} - 1\right) / 3, \text{if } a=1/\left(1+\dot{a}t\right).
\end{aligned}
\right. \label{equ::app_solution_cpaw}
\end{align}
\begin{figure}
	\includegraphics[width=\columnwidth]{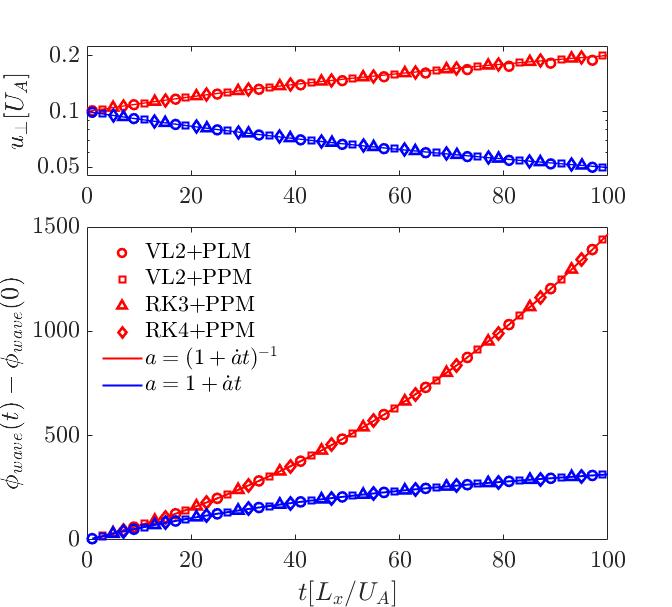}
    \caption{The evolution of amplitude and wave phase of a circularly-polarized wave in 1D expanding / compressing boxes. The box expands / shrinks with $\dot{a} L_x / U_\text{A}=0.01$. Different symbols represent different combinations of time integer and spatial reconstruction method. Red and blue refer to the case of compressing and expanding boxes, respectively. The solid lines are the analytic expectations from Equation~\ref{equ::app_solution_cpaw}. The top panel is the mean velocity amplitude $u_\perp$ and the bottom panel refers to the phase $\phi_\text{wave}(t)$.}
    \label{fig::app_cpaw_expand}
\end{figure}

In our test, a circularly-polarized Alfv\'en wave, whose wavelength is equal to the simulation box size $L_x$ resolved by 128 grid cells, moves forward along the $x$-direction initially. 
We use periodic boundaries, and the box expands / shrinks with the expansion ratio $a(t)$ set to $1+ \dot{a} t$ or $\left(1+ \dot{a} t\right)^{-1}$. The expansion or compression is quite slow compared with the wave frequency, $\dot{a}=0.01 U_A/L_x$.
We plot the amplitude and phase information $\phi_\text{wave}(t)$ in Figure~\ref{fig::app_cpaw_expand}. We apply different time integrators \footnote{Although the MHD-PIC method only supports the van Leer time integrator, we let the expanding box framework to be compatible with all available MHD time integrates.} (van Leer (VL2), third-order Runge-Kutta (RK3), and fourth-order Runge-Kutta (RK4)) and spatial reconstruction methods (piecewise linear method (PLM) and piecewise parabolic method (PPM)) using the Roe Riemann solver \citep{2020ApJS..249....4S}. The mean velocity amplitude $u_\perp$ is inversely proportional to $a$ as expected. For the wave phase $\phi_\text{wave}(t)$, we trace the peak of $u_y$ and calculate $\phi_\text{wave}(t)$ according to the peak location. In the bottom panel of Figure~\ref{fig::app_cpaw_expand}, the simulation data points closely overlap with the analytical solution (Equation~\ref{equ::app_solution_cpaw}), thus verifying the consistency and accuracy of our implementation. We also simulate waves in 2D and 3D boxes,  and the results are identical to those in the 1D simulation.

We also note that in the traditional expanding box where $a_1 = 1$, $a_2 = a_3 =a=1+\dot{a}t$, a similar solution still exists. Even though forward waves can be somehow converted to backward waves, the conversion term is largely negligible when $\dot{a}$ is much smaller then the wave frequency. In this case, as \citet{2020ApJ...891L...2S} pointed out, $w_\pm \propto a^{-0.5}$.

\subsection{Gyro-motion in expanding box}
\label{app::gyration}

\begin{figure}
	\includegraphics[width=\columnwidth]{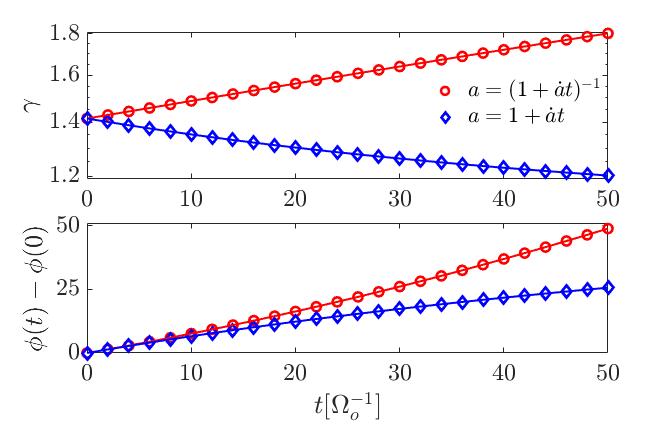}
    \caption{The time evolution of the Lorentz factor $\gamma$ and gyro phase $\phi$ of one CR in an expanding(blue) /compressing(red) box under a uniform background magnetic field. The expanding/compressing rate $\dot{a}$ is equal $0.01\Omega_0$. Markers are from simulation data and solid lines correspond to the analytic solution from Equation~\ref{equ::app_pic_expand}.}
    \label{fig::app_gyro_expand}
\end{figure}

We examine CR gyration, similar to the test in Section~\ref{sec::gyro}, but the gas is static in an expanding/compressing box. The background field $\boldsymbol{B}_g$ is still along the $x$-direction. In the case of $a_1 = 1, a_2=a_3\equiv a(t)$, the CR's equation of motion is given by,
\begin{align}
	\frac{\dd \left(a_1 p_x/m\right)}{\dd t} = 0, \notag \\
	\frac{\dd \left(a p_y/m\right)}{\dd t} = \frac{q}{mc} \frac{a p_z}{\gamma m} \frac{B_o}{a^2}, \notag \\
	\frac{\dd \left(a p_z/m\right)}{\dd t} = -\frac{q}{mc} \frac{a p_y}{\gamma m} \frac{B_o}{a^2},
\end{align}
where $B_o$ is the magnitude of initial background field $B_g\left(t=0\right)$. Clearly, the CR executes orbital motion in the $y-z$ plane, and we can express the CR momentum evolution in the form of $a p_y/m = i a p_z/m \equiv v_o \exp\left(-i \phi\left(t\right)\right)$, and $a_1 p_x / m \equiv v_\text{ini}$. Both $v_o$ and $v_\text{ini}$ are constant, given by the CR initial momentum.
For simplicity, we set $v_{\rm ini} = 0$, and the solution can be found to be (we have also tested the case with $v_{\rm ini}\neq0$ and found similar agreement)
\begin{align}
	&\gamma \left(t\right) = \sqrt{1+  v_{o}^2 / \left(  a\left(t\right)^2 \mathbb{C}^2 \right) }, \notag \\
	&\Omega_o \equiv \frac{qB_o}{mc},\notag  \\
	&\phi\left(t\right) = \frac{\Omega_o \mathbb{C}}{\dot{a}v_o} \times \notag \\ &\left\{
\begin{aligned}
& \ln \frac{v_o + \mathbb{C} \gamma\left(0\right)}{v_o/a  + \mathbb{C} \gamma\left(t\right)}, \text{if } a=1+\dot{a}t,\\
&  \frac{\mathbb{C}}{2 v_o} \left(\frac{\gamma\left(t\right)}{a} - \gamma\left(0\right)\right) - \frac{\mathbb{C}^2 }{2 v_o^2} \left(\sinh^{-1}\frac{v_o}{a\mathbb{C}} - \sinh^{-1}\frac{v_o}{\mathbb{C}} \right), \text{if } a=1/\left(1+\dot{a}t\right).
\end{aligned}
\right. \label{equ::app_pic_expand}
\end{align}
We see that $a^2 \left(p_y^2 + p_z^2\right)$ is a conserved variable, which is related to to the conservation of magnetic moment $\left(p_y^2 + p_z^2\right)/B_g = \text{Const}$.

In simulations, we choose the same parameters as in Section~\ref{sec::gyro}, $q/\left(mc\right)=B_o=\Omega_o=1$, $v_o = \mathbb{C}$ and plot the evolution of the Lorentz factor $\gamma$ and gyro phase $\phi$ in Figure~\ref{fig::app_gyro_expand}, taking $\dot{a}=0.01 \Omega_o$. The box expands/compresses much slower than the gyro frequency so that, at each moment, the CR effectively gyrates around a nearly constant magnetic field $\boldsymbol{B}_g=\boldsymbol{B}_o/a^2$.
We see that the gyro-period keeps decreasing/increasing in the expanding/compressing box, accompanied by the decrease/increase in the Lorentz factor, in perfect agreement with the analytic solution.


\bsp	
\label{lastpage}
\end{document}